\documentclass[journal]{IEEEtran} 

\usepackage{cite}

\usepackage{latexsym}
\usepackage{verbatim, amsbsy}
\usepackage{algorithm}
\usepackage{algpseudocode}

\usepackage[dvips]{graphicx}
\usepackage{url}
\usepackage{color}
\usepackage{amsmath}                      
\usepackage{amssymb}
\usepackage{amsthm}
\usepackage{amsfonts}
\usepackage{times}
\usepackage[english]{babel}
\usepackage[dvips]{graphicx}
\usepackage[normalem]{ulem}
\usepackage{bm}
\usepackage{array}
\usepackage{tikz}
\usetikzlibrary{arrows,positioning,shapes.geometric,decorations.pathmorphing,arrows.meta}
\usetikzlibrary{decorations.pathreplacing,patterns,intersections,backgrounds,calc}
\usepgflibrary{shapes.geometric}

\newcommand{\ben}{\begin{enumerate}}
\newcommand{\een}{\end{enumerate}}
\newcommand{\be}{\begin{equation}}
\newcommand{\ee}{\end{equation}}
\newcommand{\bea}{\begin{eqnarray}}
\newcommand{\eea}{\end{eqnarray}}
\newcommand{\bc}{\begin{cases}}
\newcommand{\ec}{\end{cases}}
\newcommand{\bi}{\begin{itemize}}
\newcommand{\ei}{\end{itemize}}

\newcommand{\mc}{\mathcal}

\algnewcommand{\algorithmicgoto}{\textbf{go to}}%
\algnewcommand{\Goto}[1]{\algorithmicgoto~\ref{#1}}

\algnewcommand{\algorithmicbreak}{\textbf{break}}%
\algnewcommand{\Break}[0]{\algorithmicbreak}

 \newtheorem{lem}{Lemma}
 {
 }

\def\Re{\textrm{Re}}
\def\Im{\textrm{Im}}

\def\de{\mathrm{d}}

\newcommand{\figw}{0.9\columnwidth} 

\begin{document}
\title{The Complexity-Performance Tradeoff in Resource Allocation for URLLC Exploiting Dynamic CSI}

\author{Federico~Librino,~\IEEEmembership{Member,~IEEE,}
        and~Paolo~Santi,~\IEEEmembership{Senior Member,~IEEE.}
\thanks{F. Librino is with the Italian National Research Council, 56124 Pisa, Italy (e-mail: federico.librino@iit.cnr.it).}
\thanks{P. Santi is with the Italian National Research Council, 56124 Pisa, Italy and with the Massachussets Institute of Technology, 02139 Cambridge, MA, USA (e-mail: paolo.santi@iit.cnr.it).}
\thanks{Copyright (c) 20xx IEEE. Personal use of this material is permitted. However, permission to use this material for any other purposes must be obtained from the IEEE by sending a request to pubs-permissions@ieee.org.}
}

\date{October 2020}
\maketitle

\begin{abstract}
 The challenging applications envisioned for the future Internet of Things networks are making it urgent to develop fast and scalable resource allocation algorithms able to meet the stringent reliability and latency constraints typical of the Ultra Reliable, Low Latency Communications (URLLC).
 However, there is an inherent tradeoff between complexity and performance to be addressed: sophisticated resource allocation methods providing optimized spectrum utilization are challenged by the scale of applications and the concomitant stringent latency constraints. Whether non-trivial resource allocation approaches can be successfully applied in large-scale network instances is still an open question that this paper aims to address. More specifically, we consider a scenario in which Channel State Information (CSI) is used to improve spectrum allocation in a radio environment that experiences channel time correlation.
 Channel correlation allows the usage of CSI for longer time before an update, thus lowering the overhead burden. Following this intuition, we propose a dynamic pilot transmission allocation scheme in order to adaptively tune the CSI age.
 We systematically analyze the improvement of this approach applied to a sophisticated, recently introduced graph-based resource allocation method that we extend here to account for CSI.
 The results show that, even in very dense networks and accounting for the higher computational time of the graph-based approach, this algorithm is able to improve spectrum efficiency by over 12\% as compared to a greedy heuristic, and that dynamic pilot transmissions allocation can further boost its performance in terms of fairness, while concomitantly further increase spectrum efficiency of 3-5\%.
\end{abstract}

\section{Introduction}
In the next future, the foreseeable spread of the Internet of Things technologies is expected to highly contribute to the development of the Industry 4.0 paradigm.
Coping with the stringent requirements of Industrial Internet of Things (IIoT) is extremely challenging, since high precision and low latency are both fundamental for the vast majority of applications and scenarios. In addition, dense networks, involving tens or hundreds of nodes, are expected to be deployed in order to accomplish complex tasks and coordinate multiple systems and interfaces.

In order to match the required performance, communications among huge numbers of operating devices must rely on suitable strategies and protocols.
To this aim, Ultra-Reliable Low-Latency Communications (URLLC) are currently being extensively investigated, and are seen as a promising approach to tackle the issues posed by highly demanding scenarios.
URLLC, and their application to IIoT, are in fact a fundamental brick of the 5G and emerging 6G systems, whose vision encompasses, among the others, the development of massive machine-to-machine communications~\cite{Rew1in,Rew2in}.
To this aim, URLLC management needs innovative approaches, since both a very high reliability (with an error rate of $10^{-5}$ or even lower) and a very stringent latency constraint, in the range of a few milliseconds, are enforced.
In order to efficiently utilize the available spectrum, orthogonal resource allocation through OFDMA (Orthogonal Frequency Division Multiple Access) is commonly regarded as a feasible technique, due to its robustness and flexibility. One of the main issues is hence how to allocate the available time-frequency resources. Distributed solutions are generally simpler to implement, but the lack of a coordination entity makes it necessary to introduce redundancy with the aim of correcting the errors due to unwanted collisions among transmissions.

Conversely, centralized allocations can be optimized through properly designed algorithms, provided that all the necessary information is available at the central entity.
This poses a strong limitation when networks scale to large numbers of devices, since optimized allocations usually require a heavy (non polynomial) computational burden, which is hardly compatible with the stringent timing constraints of URLLC.
On the opposite side, much faster heuristics have been also developed. While they offer the advantage of a quick computation (even linear with the number of transmitters), as the network grows they can easily fall far off the actual optimum and lead to substantial resource waste.

In our previous work~\cite{OurIoT}, we proposed an \emph{in between} solution: an algorithm based on graph theory which was proved to run in polynomial time, yet attaining high performance in terms of both served traffic and fairness.
Although promising, the obtained results were limited to a static scenario, where Channel State Information (CSI) is not available and allocation is performed based only on statistical knowledge of the channel conditions. However, a highly dynamic radio environment is expected in most Industry 4.0 applications, and whether non-trivial optimization algorithms can be successfully used in this context as well is an open question that this paper seeks to address.

The need for a more efficient usage of the available spectrum in a dynamic radio environement makes it relevant to acquire and exploit updated information about the channel condition.
The collection and exploitation of CSI raises two additional issues.
Firstly, CSI must be conveyed to the central entity, meaning that a fraction of the resources must be devoted to pilot transmissions. The exact amount of such overhead resources must be carefully planned: increasing it improves the accuracy of the CSI, but reduces the spectrum left for data transmission; lowering it grants more resources for data, making it easier to match the reliability and latency constraints, but allocation must rely on less precise CSI.
From this viewpoint, channel time correlation, which generally impairs the communications by limiting time diversity, can instead be helpful, since CSI needs to be updated less frequently.
Secondly, as the CSI is collected and refreshed, resource allocation should be recomputed so as to follow the changes in the channel conditions, even for applications with deterministic and periodic traffic patterns. This also requires time, depending on the utilized algorithm, and in fast changing scenarios the obtained allocation solution might be already too old to achieve satisfying performance.

In this work, we investigate these aspects in an uplink industrial scenario, with hundreds of sensors collecting information and transmitting it to a single Access Point (AP). The scenario may represent a smart factory, where a discrete automation application is considered. As in~\cite{M5}, transmissions are considered to be periodic, and a semipersistent cyclic reservation is needed.
We consider a realistic scenario, where the spectrum is intensively reused among several coexisting networks. This reflects into two main consequences. The available spectrum might be fragmented into non adjacent sub-bands, thus implying that different frequency channels experience different channel conditions. Besides, even if proper frequency reuse techniques are employed in order to mitigate the inter-network interference, this cannot be completely avoided, and different channels might be differently affected by residual interference.
All these factors make an efficient resource allocation an even harder problem, and a timely exploitation of the CSI pivotal.

\subsection{Related Work}
The very tight latency constraint required for URLLC has led the 3rd Generation Partnership Project to suggest proposals for a novel frame structure for OFDMA systems~\cite{M23}, with slots shorter than 1 ms.
Several resource allocation schemes, able to attain good performance in this type of scenario, have been then investigated~\cite{M3,M4,M5,M6,M25,M27}.
Distributed schemes avoid the burden of a centralized computation and the need for a resource reservation phase. Hence, this approach leads to much faster algorithms, at the cost of allowing potentially interfering communications. When dealing with very dense networks, this may stronlgy impair system reliability, since retransmissions would make it difficult to match the stringent latency constraint.
Conversely, centralized schemes can ensure orthogonal resource allocation, avoiding errors due to collisions among simultaneous transmissions. The drawback lies in the need for an efficient centralized allocation schemes.
Authors in~\cite{M1} address the problem of a real-time wireless control system based upon URLLC, and propose a suboptimal solution aiming at maximizing the uplink spectral efficiency. In\cite{M5}, two scenarios are compared: a centralized periodic reservation is proposed for deterministic packet arrivals, similar to our work, while a grant-free distributed access to a common pool of resources is outlined for sporadic packet arrivals.
Centralized resource allocation based on the solution of a mixed integer non convex optimization problem is presented in~\cite{M8}, where power allocation is also tackled. An approximated solution requiring polynomial time is derived, and applied to the downlink, but the number of users remains much lower than in our work.
In addition to spectrum and power allocation, the optimal antenna configuration in a MIMO (Multiple-Input-Multiple-Output) scenario is also sought in~\cite{M11}, with the aim of maximizing energy efficiency.
Spatial diversity is shown to be unable to ensure both very high reliability and very low latency with finite power in~\cite{M13}: here, queueing delay is also considered, and a cross layer optimization method is detailed.
The application of URLLC to device-to-device (D2D) communications in envisioned in~\cite{M15}, where authors derive an optimal power control to maximize the average D2D rate, with a constraint on the outage probability at the Base Station.
Recently, tools from machine learning (ML) have been also considered for solving this and other similar types of allocation problems~\cite{ML1,ML2,ML3}. Due to its inherent adaptability, this approach is particularly effective when the correlations between channel measurements, energy efficiency, environmental parameters and transmissions reliability are too complex to be modeled in a tractable way. In~\cite{ML1}, authors use deep reinforcement learning to optimize bandwidth allocation between URLLC and enhanced Mobile Broad Band services targeting a system wide tradeoff.
Adaptive channel assignment under the cognitive communication paradigm is instead investigated in~\cite{ML2}, where fountain codes are used to achieve the target reliability, while ML is exploited to match the latency constraint.
In most works, CSI is assumed to be perfect and readily available~\cite{M11,M13,M15}. The impact of imperfect CSI is instead quantified in~\cite{M26,M27,M28,M29}. In~\cite{M26}, resource allocation includes also the assignment of resources to pilot signaling and feedback overhead, but the scenario is not time varying, and the resulting non-linear integer programming problem is solved via exhaustive search over a small feasible solution space.
Imperfect CSI only is assumed in~\cite{M27}, with a bounded channel error, and an iterative algorithm is outlined, but for low numbers of active devices.
The issue of imperfect CSI is also tackled in~\cite{M28}, where massive Multi User MIMO deployment is analyzed. The optimal pilot length is attained through linear minimum error probability detection, and the optimal fraction of resources reserved to pilot signal is found to be around 30\% in this specific setting. A similar result is attained also in~\cite{M29}, where the authors use instead the golden section search method to quickly find the optimal pilot length.
Finally, \cite{M30} addresses the coexistence of URLLC and enhanced Massive Broad Band through network slicing. The Max-Matching Diversity channel allocationm, originally proposed in~\cite{M31}, is utilized to minimize the number of users in outage. In our work, the adopted algorithm also relies on maximum matching over a bipartite graph, but weights are assigned to edges in order to incorporate the latency constraints into the resource allocation algorithm, while still keeping a polynomial complexity.

\subsection{Our contribution}
In our work, we aim at investigating the tradeoff between CSI accuracy and data transmission reliability from a novel perspective. In most works, part of the spectrum recources are reserved to pilot signaling, which is then immediately exploited to enhance decoding reliability over a predetermined set of allocated resources.
However, due to the very short time scale of URLLC, channel time correlation cannot be neglected. While this aspect often impairs systems based on retransmissions, due to reduced time diversity, we instead exploit this correlation in order to limit the overhead required for CSI.
Actually, channel time correlation, especially if strong, makes channel estimations useful for longer times, and channel measurements can still be used for subsequent resource allocation procedures, although with lower effectiveness, as time passes.
In our approach, CSI is hence used proactively, together with topology and interference knowldege, in order to derive the amount of resources required for each transmitting device in order to reach the target reliability.
Aiming at improving fairness, we also introduce a smart and dynamic pilot transmissions allocation, since differences in link lengths and measured fading coefficients can be better exploited with a tunable CSI update frequency.
Since information age is a key issue for a resource allocation based on already stored channel measurements, the computational time of a resource allocation algorithm also plays a pivotal role. Algorithms taking longer times make the usage of stored CSI less effective, and this may become dramatic in very dense networks, with hundreds of devices.
We hence analyze the GBA (Graph Based Algorithm) allocation algorithm outlined in~\cite{OurIoT}, which was proved to run in polynomial time, in order to assess its capability of leveraging the channel information provided by the transmitters.

Summarizing, the original contribution of this article are the following:
\begin{enumerate}
 \item We introduce the notion of dynamic CSI in the context of URLLC, where flexible allocation of pilot resources is done both in terms of frequency of CSI update, as well as in expected utility of the update to increase fairness in channel access;
 \item We modify the GBA resource allocation algorithm in order to include an in-band pilot transmission from devices so as to provide channel estimation at the AP and to exploit it;
 \item We analyze the tradeoff between CSI update frequency and overall system performance in terms of served traffic, in a dense network scenario;
 \item We assess the impact of the algorithm computational time on the overall performance, comparing GBA with a greedy heuristic algorithm running in quasi-linear time;
 \item We evaluate the benefits of dynamic pilot transmission allocation in terms of both system fairness and spectrum efficiency.
\end{enumerate}
We organize the paper material as follows. In Section~\ref{sec:GBA}, we briefly describe the analyzed scenario and the graph-based allocation algorithm GBA. We also highlight the modifications required for GBA to include also pilot signaling for CSI estimation. Section~\ref{sec:sysmod} details the time correlated channel model, as well as how the pilot signaling is organized, and the dynamic pilot transmissions allocation scheme is outlined in Section~\ref{sec:dynpilot}.
The obtained results are presented in Section~\ref{sec:results}, while Section~\ref{sec:conclu} concludes the paper.

\section{Analyzed Scenario and GBA Overview}
\label{sec:GBA}
We consider a scenario where $N$ devices are randomly placed in a circular area of radius $L$ around a common Access Point (AP). Each device generates a data packet of size $\ell$ bits periodically, with period $\nu$ ms, and must convey the packet to the AP with reliability $\rho$ and within a predefined delay, otherwise the packet is considered to be lost.

The available spectrum must be shared among all devices, and is reused through an OFDMA (Orthogonal Frequency Division Multiple Access) based approach. To this aim, time is divided into slots of duration $\tau$ ms, corresponding to $n_t$ subsequent OFDM symbols. Hence, we can define a \emph{cycle} as a sequence of $T = \nu/\tau$ time slots, with each device $D_i$ generating one packet per cycle at a fixed issue time $t_i\in\{1,2,\ldots,T\}$.
The delay constraint is $\Delta$ slots, meaning that the packet from user $D_i$ must be delivered within time slot $t_i+\Delta$. Different packet delays for different devices can be considered as well, but this does not alter the allocation procedure, hence we opt for a single value of $\Delta$.

We assume that $C$ frequency channels are available, each composed of $n_c$ adjacent subcarriers. Channels need not be contiguous in the spectrum. We define a Resource Unit (RU) as the basic time-frequency resource unit, which spans over one frequency channel and one time slot. Within each cycle, thus, $TC$ RUs are available to be assigned to (some of) the $N$ devices.

Part of the RU can be reserved for pilot transmission, in order to acquire CSI and the AP. We call them \emph{Pilot RUs} (pRUs), and they are generally fixed and known to all the devices. The amount of these pRUs and how they are exploited depends on the specific allocation algorithm, and is described in the next sections.

The allocation of the resources is performed in a centralized manner by the AP through the GBA algorithm introduced in~\cite{OurIoT} and whose functioning is recalled in the next section. The information required to run the algorithm are collected at the AP, which then broadcasts the overall assignment over a dedicated control downlink channel, not modeled in this work.

\subsection{The Graph-Based Allocation Algorithm (GBA)}
In this section, we briefly summarize the GBA algorithm, and outline the minor modifications required to reserve some of the RUs to pilot signaling.

The GBA algorithm allocates the highest fraction of the $N$ devices on the $CT$ RUs. Every device can be assigned a subset of subsequent RUs on the same frequency channel, whose number depends on the device and on the selected channel as well.
More precisely, we define as $\mathcal{F}(c,i,\rho)$ the number of RUs required on channel $c$ in order for device $D_i$ to transmit with reliability $\rho$.
The exact expression of this function depends on the application scenario, and is computed, for all the devices and the channels, at the beginning of the GBA execution. Furthermore, a set of $C$ variables $\beta_1,\beta_2,\ldots,\beta_C$ is also initialized, where $\beta_c\in\{0,1,\ldots,T\}$ indicates the latest already allocated RU on channel $c$. At the beginning all the RUs are free, hence all the $\beta_c$'s are set to 0.

The GBA works in subsequent iterations. At the beginning of each iteration, GBA creates a weighted bipartite graph between two sets of nodes, $\mathcal{C}$ and $\mathcal{N}$. The former is the set of the $C$ available channels, while the latter is the set of the $N$ devices yet to be allocated.
An edge $(X_c,Y_i)$ between node $X_c\in\mathcal{C}$ corresponding to channel $c$, and node $Y_i$ corresponding to device $D_i$ exists only if the following condition holds:
\begin{equation}
 \mathcal{F}(c,i,\rho) \leq \Delta.
\end{equation}
The condition ensures that the number of RUs needed for a reliable trasmission from $D_i$ to the AP on channel $c$ is lower than the maximum allowed latency, hence channel $c$ is a feasible channel for $D_i$. The weight of this edge is given by
\begin{equation}
 w(c,i) = T+\Delta -\left(\max(\beta_c,t_i-1) + \mathcal{F}(c,i,\rho)\right),
 \label{weights}
\end{equation}
Looking at equation (\ref{weights}), the term $\max(\beta_c,t_i-1)+1$ indicates the first free RU that would be allocated to device $D_i$ if $D_i$ were assigned to this channel. Since $\mathcal{F}(c,i,\rho)$ RUs are then necessary to $D_i$ for a reliable transmission on channel $c$, the term $\left(\max(\beta_c,t_i-1) + \mathcal{F}(c,i,\rho)\right)$ corresponds to the last RU that would be reserved to $D_i$.
Given that $T$ RUs are available on each channel, the weight in (\ref{weights}) simply corresponds to the number of remaining free slots on channel $c$ after allocating device $D_i$ to it, plus an offset $\Delta$.
After having evaluated the existence and weight of all the edges, the Hungarian algorithm~\cite{hungarian} is run to find the maximum weighted matching over this graph.
If edge $(X_c, Y_i)$ belongs to the matching, then the RUs on channel $c$ from slot $\max(\beta_c,t_i-1)+1$ to slot $\max(\beta_c,t_i-1)+\mathcal{F}(c,i,\rho)$ are assigned to device $D_i$.
The weight expression in (\ref{weights}) ensures that allocation is made in such a way that the quantity of still available RUs after this phase is maximized.
For any channel $c$ on which a device $D_i$ has been assigned, the corresponding variable $\beta_c$ is updated to $\max(t_i-1, \beta_c)+\mathcal{F}(c,i,\rho)$, thus still containing the index of the last busy RU on this channel.
At most $C$ out of the $N$ devices are allocated in the first phase. These devices are removed from the set $\mathcal{N}$, and the second phase begins.
A new bipartite graph is built between $\mathcal{C}$ and the updated set $\mathcal{N}$. The existing condition for edge $(X_c,Y_i)$ is now slightly different to account for the already allocated RUs, and becomes
\begin{equation}
 \max(\beta_c,t_i-1) + \mathcal{F}(c,i,\rho) < t_i + \Delta.
 \label{edgecond}
\end{equation}
In addition, all the nodes in $\mathcal{N}$ with degree 0 in this new bipartite graph are also excluded from $\mathcal{N}$: the corresponding devices cannot be served in this allocation round without violating either the reliability or the latency constraint.
From this point on, the phase proceeds exactly as the first one: the weights, which might be different, due to the modified values of the $\beta_c$'s, are assigned to the edges, and a new maximum weighted matching is derived.
At each subsequent phase, at most $C$ devices are allocated and removed from $\mathcal{N}$. The algorithm proceeds until the set $\mathcal{N}$ is empty, meaning that all the devices have been either allocated or excluded.

\subsection{Integration of pilot transmission}
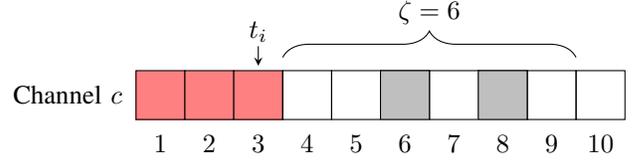
\begin{figure}
 \centering
 \begin{tikzpicture}[>=stealth,scale=0.65]
  \foreach \x in {0,1,...,2}
      \filldraw[color=red!50, draw=black] (\x, 0) rectangle +(1,1) ;
  \foreach \x in {3,4,...,9}
      \draw[draw=black] (\x, 0) rectangle +(1,1);
      
  \foreach \x in {1,2,...,10}
      \node at (\x-0.5, -0.5) {$\x$};
      
  \filldraw[color=gray!50, draw=black](5,0) rectangle +(1,1);
  \filldraw[color=gray!50, draw=black](7,0) rectangle +(1,1);
  
  \node at (2.5, 1.8) {$t_i$};
  \draw[->] (2.5, 1.5) -- (2.5, 1.1);
  
  \node at (-1.4, 0.5) {Channel $c$};
  
  \draw[decorate,decoration={brace,amplitude=10pt},xshift=0pt,yshift=2pt] (3,1.2) -- +(6,0) node[black,midway,yshift=0.6cm]{$\zeta=6$};
 \end{tikzpicture}
 \caption{Example of resource allocation on a channel partially used for pilot transmission. Red squares are already allocated RUs, hence $\beta_c=3$ for this channel. Grey squares are the pRUs reserved to pilot signaling, and cannot be allocated. We assume that for user $D_i$ the number of required RUs is $\mathcal{F}(c,i,\rho)=4$, and that the data packet is issued at time slot $t_i=3$. Hence, the RUs up to slot 9 are assigned to $D_i$, with $\zeta=6$.}
 \label{fig:allofeed}
\end{figure}

The GBA functioning illustrated in the previous section assumes that all the $CT$ RUs in a cycle are free for data transmission. This may not be the case, if some of them are instead pRUs reserved to pilot transmission.
We now illustrate how GBA is to be slightly modified to take this issue into account. We assume that the pRUs are known and fixed.
In each iteration of the algorithm, when a new bipartite graph is built, the weight of edge $(c,i)$ must consider the pRUs on channel $c$, which may increase the overall transmission time. The modified formulation of the edge existence condition reads as
\begin{equation}
 \max(\beta_c,t_i-1) + \zeta < t_i + \Delta,
 \label{modicond}
\end{equation}
being $\zeta$ the minimum value such that the number of available RUs on channel $c$ from slot $\max(\beta_c,t_i-1)+1$ to slot $\max(\beta_c,t_i-1)+\zeta$ (included) is equal to $\mathcal{F}(c,i,\rho)$. This is equivalent to computing the number of required RUs for the device transmission, starting from slot $\max(\beta_c,t_i-1)+1$ and \emph{jumping} the pRUs until $\mathcal{F}(c,i,\rho)$ are allocated to $D_i$, as illustrated in Figure \ref{fig:allofeed}.
Similarly, the weight assigned to the edge to compute the matching is now expressed as
\begin{equation}
 w(c,i) = T+\Delta -\left(\max(\beta_c,t_i-1) + \zeta\right),
 \label{modweights}
\end{equation}
while the other aspects of the algorithm remain unaltered.

\section{System Model}
\label{sec:sysmod}
\subsection{Correlated Channel Model}
The wireless transmission from a device to the AP is affected by both path loss and fading. The former depends on the physical distance between transmitter and receiver, while the latter is modeled as a complex Gaussian random variable with zero mean and unit variance. The fading coefficient of each channel is considered constant across the entire cycle, and independent across channels.
Hence, the received Signal to Interference plus Noise Ratio (SINR) from user $D_i$ on channel $c$ at cycle $m$ is expressed as
\begin{equation}
 \Gamma_{i,c,m} = \frac{P_Td_i^{-\alpha}}{N_0+I_c}|h_{i,c,m}|^2 = \frac{\Gamma_Td_i^{-\alpha}}{\Lambda_c}|h_{i,c,m}|^2,
 \label{SINR}
\end{equation}
where $\alpha$ is the path loss exponent, $d_i$ is the distance between $D_i$ and the AP, $P_T$ is the transmit power, which is the same for all the devices, $I_c$ is the interference on channel $c$ coming from surrounding transmissions, $N_0$ is the noise power and $h_{i,c,m}$ is the fading coefficient.
We assume that the resources are orthogonally allocated, meaning that there is no interference among users. However, in a realistic scenario the spectrum is often intensively reused, and it is unlikely that interference coming from surrounding systems can be completely removed.
In our work, we consider $I_c$ as a residual interference and model it as additional noise $Y_cN_0$, where $Y_c\sim\mathcal{U}(0,Y_M)$ is a real uniform random variable. The residual interference level is considered to be slowly varying over time, and is measured at the AP at the beginning of each cycle. In (\ref{SINR}), we also define the transmit SNR as $\Gamma_T=P_T/N_0$, and $\Lambda_c=1+Y_c$ is an equivalent coefficient to measure the channel interference level.
Given the random distribution of the fading coefficient, its squared value $|h_{i,c,m}|^2$ is an exponential random variable with parameter 1.

Due to the very short duration of a cycle, temporal correlation should not be neglected. Hence, we consider a temporally correlated fading, modeled through a first-order Gauss-Markov process~\cite{New1} with correlation factor $\gamma$ across subsequent cycles, such that
\begin{equation}
 h_{i,c,m+1} = \gamma h_{i,c,m} + \sqrt{1-\gamma^2}\xi,
 \label{corrchan}
\end{equation}
where $\xi$ is also a complex Gaussian random variable with zero mean and unit variance.

\subsection{Derivation of $\mathcal{F}(c,i,\rho)$}
\begin{figure}
 \begin{center}    
 \includegraphics[width=\figw]{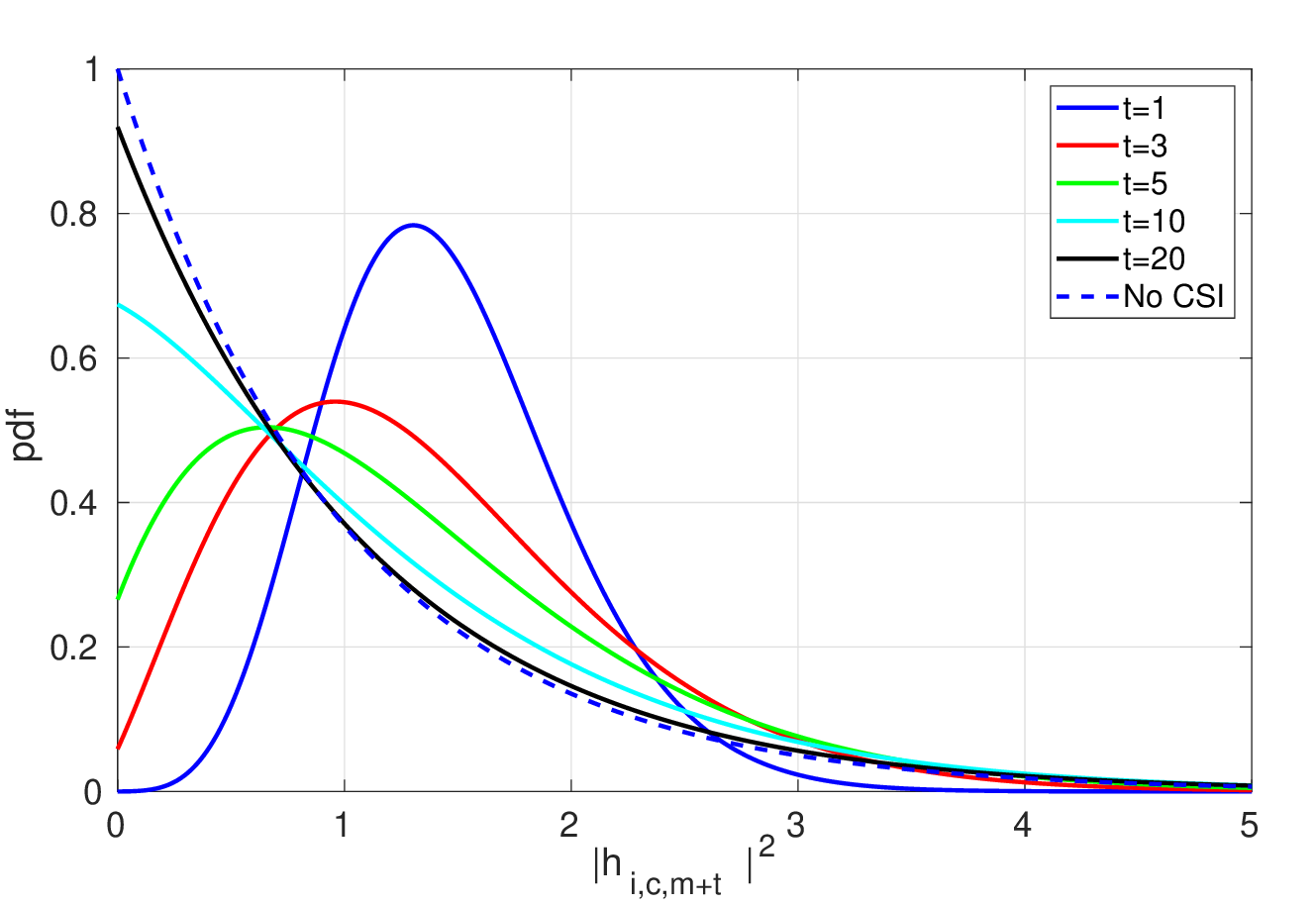}
 \caption{\small Conditional pdf of $|h_{i,c,m+t}|^2$ when $|h_{i,c,m}|^2=1.5$, for different values of the temporal distance $t$.}
 \label{fig:FB_corrchan}
 \vspace{-0.7cm}
 \end{center}
\end{figure}

The allocation algorithm GBA, described in Section~\ref{sec:GBA}, prescribes that each device can be assigned RUs on the same channel, over which a data packet is transmitted. The data packet is correctly received if all its parts, sent over different RUs, have been correctly decoded. Since the fading coefficient is considered fixed over the whole cycle, the $\ell$ bits of the data packet are assumed to be equally spread over all the $k$ assigned RUs, with a resulting rate of $\ell/(k\tau)$ bit/s.
We derive the decoding probability $p_{\textrm{dec}}$ by using the Shannon approximation\footnote{The Short Packet Communication (SPC) model~\cite{M12}, which is more suitable for very short transmissions, as well as the approximation based on linearization in~\cite{Zorzi}, are shown in~\cite{OurIoT} to be well approximated by the Shannon expression in this type of scenario.}: an outage occurs when the resulting transmission rate exceeds the instantaneous channel capacity $\Omega = B\log_2(1+\Gamma_{i,c,m})$, where $B$ is the channel bandwidth.
Hence, we get
\begin{eqnarray}
 p_{\textrm{dec}} & = & \mathbb{P}\left[B\log_2(1+\Gamma_{i,c,m}) \geq \frac{\ell}{k\tau}\right] \nonumber \\
 & = & \mathbb{P}\left[|h_{i,c,m}|^2 \geq \frac{\Lambda_c}{\Gamma_Td_i^{-\alpha}}\left(2^{\frac{\ell}{kq}}-1\right)\right]
\end{eqnarray}
where $q=B\tau$.
In order to match the reliability constraint, we must ensure that $p_{\textrm{dec}}\geq\rho$ by properly tuning the number $k$ of assigned RUs. When no CSI information is available, $|h_{i,c,m}|^2$ is exponentially distributed, and the minimum number $\mathcal{F}(c,i,\rho)$ of RUs to be allocated to device $D_i$ on channel $c$ is~\cite{OurIoT}
\begin{equation}
 \mathcal{F}(c,i,\rho) = \left\lceil\frac{\ell}{q }\left[\log_2\left(1-\frac{\Gamma_T\ln(\rho)}{\Lambda_c d_i^{\alpha}}\right)\right]^{-1}\right\rceil.
 \label{deffi}
\end{equation}
However, due to the channel temporal correlation, the knowledge of the fading coefficient $|h_{i,c,m}|^2$ at a given cycle $m$ modifies the distribution of $|h_{i,c,m+t}|^2$ according to the following Lemma:
\begin{lem}
 The probability distribution function $f_t(x|z)$ of $|h_{i,c,m+t}|^2$ given that $|h_{i,c,m}|^2=z$ can be expressed as
 \begin{eqnarray}
 f_t(x|z) & = & \mathbb{P}[\left.|h_{i,c,m+t}|^2=x\right||h_{i,c,m}|^2=z] \nonumber \\
 & = & \frac{1}{b}e^{-\frac{x}{b}}e^{-\frac{a^2z}{b}}I_0\left(\frac{2a}{b}\sqrt{xz}\right),
 \label{corrpdf}
\end{eqnarray}
where
\begin{equation}
 a = \gamma^t, \quad \quad b=(1-\gamma^2)\sum_{j=0}^{t-1}\gamma^{2j},
\end{equation}
while $I_0(\cdot)$ is the modified Bessel function of the first kind of order 0.
\label{lem:corrchan}
\end{lem}
\begin{proof}
 See Appendix~\ref{app:proof2}.
\end{proof}
\vspace{-.5cm}
By looking at the pdf in (\ref{corrpdf}), it can be observed that $a\rightarrow 0$ and $b\rightarrow 1$ as $t\rightarrow+\infty$; hence, the pdf tends to an exponential distribution with unitary mean, which is correct, since this case corresponds to having no CSI at all. As an example, in Figure~\ref{fig:FB_corrchan} we show the pdf of $|h_{i,c,m+t}|^2$ when $|h_{i,c,m}|^2$ is equal to 1.5, for different values of $t$.
As $t$ increases, not only does the variance of the distribution increase, but also the probability of very low values becomes higher, thus making it necessary to allocate more RUs in order to match the high reliability constraint.
This confirms the importance of accessing recent and relevant CSI in order to reduce the amount of necessary resources and thus have room for a higher number of devices.

In case of channel correlation, the number $\mathcal{F}(c,i,\rho,t|z)$ of RUs required for device $D_i$ on channel $c$ at cycle $m$ given the known value
$|h_{i,c,m-t}|^2 = z$ is the value $k$ that gives $p_{\textrm{dec}}=\rho$, which in this case reads
\begin{equation}
 \mathbb{P}\left[\left.|h_{i,c,m}|^2 \geq \frac{\Lambda_c}{\Gamma_Td_i^{-\alpha}}\left(2^{\frac{\ell}{kq}}-1\right)\right| |h_{c,i,m-t}|^2=z\right] = \rho.
 \label{condpro}
\end{equation}
By calling $F_t(x|z)=\int_0^xf_t(u|z)\,\de u$ the conditional cumulative distribution function (CDF) of the fading coefficient, we can elaborate (\ref{condpro}) into
\begin{eqnarray}
 F_t\left(\left.\frac{\Lambda_c}{\Gamma_td_i^{-\alpha}}\left(2^{\frac{\ell}{kq}}-1\right)\right|z\right) & = & 1 - \rho \nonumber \\
 \frac{\Lambda_c}{\Gamma_td_i^{-\alpha}}\left(2^{\frac{\ell}{kq}}-1\right) & = & F_t^{(-1)}\left(1-\rho|z\right),
\end{eqnarray}
which finally yields
\begin{equation}
 \mathcal{F}(c,i,\rho,t|z) = \left\lceil\frac{\ell}{q}\left[\log_2\left(\!1+\frac{\Gamma_T}{\Lambda_cd_i^{\alpha}}F_t^{(-1)}(1-\rho|z)\right)\right]^{-1}\right\rceil.
 \label{deffires}
\end{equation}
Equation (\ref{deffires}) must be solved numerically, due to the non closed form of $F_t(x|z)$. Nevertheless, when implementing the algorithm into a real system, a fast solution can be obtained by quantizing the possible values of the channel fading coefficient and building a matrix $\mathbf{F}$ offline. Element $\mathbf{F}[t,r]$ contains the value $F_t^{(-1)}(1-\rho|z)$, being $z$ the midpoint of the $r$--th quantization bin of the fading coefficient.
This matrix is hence checked to quickly derive the number of required RUs through (\ref{deffires}). Notice that as $t$ grows, the distribution $F_t(x|z)$ tends to the CDF an exponential distribution, namely $1-e^{-x}$, thus leading $\mathcal{F}(c,i,\rho,t|z)$ back to the expression of $\mathcal{F}(c,i,\rho)$ in (\ref{deffi}).

\begin{figure}
 \begin{center}    
 \includegraphics[width=\figw]{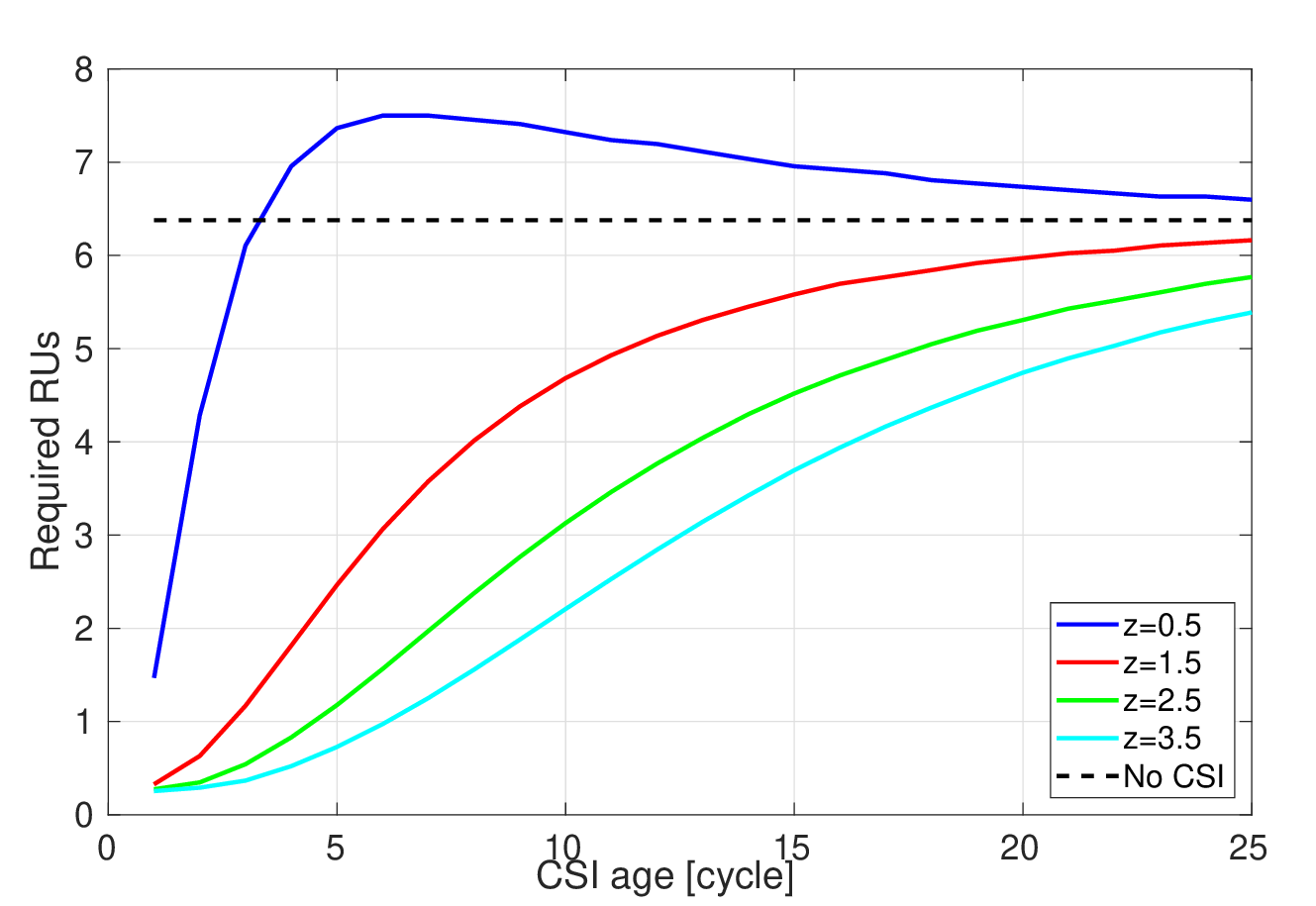}
 \caption{\small Needed resources as a function of the CSI age, for different values of the measured channel fading coefficient. Here the device is placed at a distance $d=40$ m from the AP, and the parameters in Table~\ref{tab:Sympara} are used.}
 \label{fig:FB_RU_vs_CSIage}
 \vspace{-0.3cm}
 \end{center}
\end{figure}
In Figure~\ref{fig:FB_RU_vs_CSIage}, we highlight the importance of a timely CSI in terms of needed RUs in order to match the reliability constraint. We plot the non rounded value given by (\ref{deffi}) and (\ref{deffires}) as a function of the CSI age $t$ measured in cycles, for $\gamma=0.95$. It is immediately seen that, when the measured CSI $z$ at a given cycle $m$ is high, having this information can lead to dramatic decrease in the number of needed RUs at cycle $m+t$.
Even when $t$ is large (e.g., $t=10$), less than half of the RUs are enough, if the measured coefficient $z$ was high. Things are different if instead the measured channel was experiencing harsh conditions ($z<1$). In this case, CSI is still useful, but only if it is fresh ($t<4$ cycles). In this case, in fact, having an imperfect or outdated knowledge (hence with larger variance) of a bad channel results in requiring more RUs than having no knowledge at all.

\subsection{Providing Channel State Information}
In order to obtain CSI, part of the $CT$ RUs in each cycle need to be reserved for pilot transmission. In this paper, we assume than perfect CSI of a channel can be obtained through a pilot signal that extends over a single RU on that channel. While this is a reasonable assumption in this specific scenario~\cite{CSIerr}, considering estimation errors and the tradeoff between its magnitude and the pilot length is left as an interesting future extension.

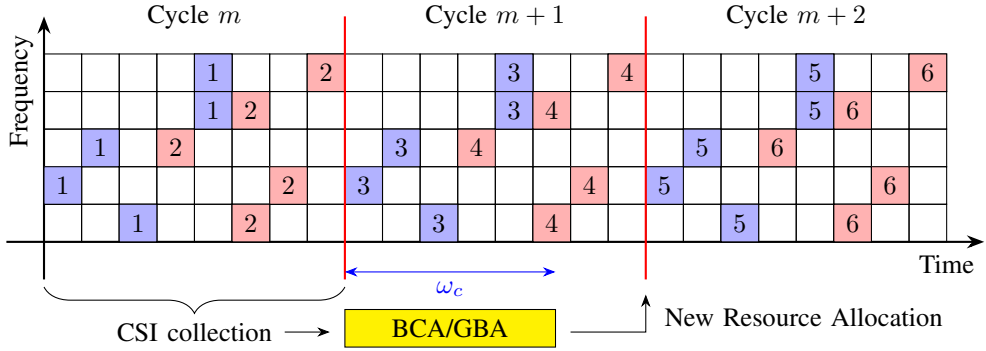
\begin{figure*}[t]
 \centering
 \begin{tikzpicture}[>=Stealth]
    \tikzset{oru/.style={draw=black,shape=rectangle,fill=blue!30,minimum height=0.5cm,minimum width=0.5cm, inner sep = 0.01cm},
    aru/.style={draw=black,shape=rectangle,fill=red!30,minimum height=0.5cm,minimum width=0.5cm, inner sep = 0.01cm}
    }
    \def\uni{0.5}; 
    \def\til{0.25}; 
    \def\cyc{4}; 
    \foreach \x in {0,1,...,23}
     \foreach \y in {0,1,...,4}
      \draw[draw=black] (\x/2, \y/2) rectangle +(\uni,\uni);

    \node at (2, 3) {Cycle $m$};
    \node at (6, 3) {Cycle $m+1$};
    \node at (10, 3) {Cycle $m+2$};
    
    \draw[->,thick] (-0.5,0) -- (12.5,0);
    \node at (12,-0.3) {Time};
    
    \draw[->,thick] (0,-0.5) -- (0,3);
    \node at (-0.3,2) [rotate=90] {Frequency};

    \node at (3*\uni-\til,1*\uni-\til) [oru] {$1$};
    \node at (1*\uni-\til,2*\uni-\til) [oru] {$1$};
    \node at (2*\uni-\til,3*\uni-\til) [oru] {$1$};    
    \node at (5*\uni-\til,4*\uni-\til) [oru] {$1$};
    \node at (5*\uni-\til,5*\uni-\til) [oru] {$1$};
    
    \node at (6*\uni-\til,1*\uni-\til) [aru] {$2$};
    \node at (7*\uni-\til,2*\uni-\til) [aru] {$2$};
    \node at (4*\uni-\til,3*\uni-\til) [aru] {$2$};    
    \node at (6*\uni-\til,4*\uni-\til) [aru] {$2$};
    \node at (8*\uni-\til,5*\uni-\til) [aru] {$2$};
    
    \begin{scope}[xshift=\cyc cm]
     \node at (3*\uni-\til,1*\uni-\til) [oru] {$3$};
     \node at (1*\uni-\til,2*\uni-\til) [oru] {$3$};
     \node at (2*\uni-\til,3*\uni-\til) [oru] {$3$};     
     \node at (5*\uni-\til,4*\uni-\til) [oru] {$3$};
     \node at (5*\uni-\til,5*\uni-\til) [oru] {$3$};
     
     \node at (6*\uni-\til,1*\uni-\til) [aru] {$4$};
     \node at (7*\uni-\til,2*\uni-\til) [aru] {$4$};
     \node at (4*\uni-\til,3*\uni-\til) [aru] {$4$};    
     \node at (6*\uni-\til,4*\uni-\til) [aru] {$4$};
     \node at (8*\uni-\til,5*\uni-\til) [aru] {$4$};
    \end{scope}
    
    \begin{scope}[xshift=2*\cyc cm]
     \node at (3*\uni-\til,1*\uni-\til) [oru] {$5$};
     \node at (1*\uni-\til,2*\uni-\til) [oru] {$5$};
     \node at (2*\uni-\til,3*\uni-\til) [oru] {$5$};     
     \node at (5*\uni-\til,4*\uni-\til) [oru] {$5$};
     \node at (5*\uni-\til,5*\uni-\til) [oru] {$5$};
     
     \node at (6*\uni-\til,1*\uni-\til) [aru] {$6$};
     \node at (7*\uni-\til,2*\uni-\til) [aru] {$6$};
     \node at (4*\uni-\til,3*\uni-\til) [aru] {$6$};    
     \node at (6*\uni-\til,4*\uni-\til) [aru] {$6$};
     \node at (8*\uni-\til,5*\uni-\til) [aru] {$6$};
    \end{scope}
    
    \draw[thick,red] (4,-0.5) -- (4,3);
    \draw[thick,red] (8,-0.5) -- (8,3);
    
    \draw[decorate,decoration={brace,amplitude=10pt},xshift=0pt,yshift=0pt] (\cyc,-0.6) -- +(-\cyc,0) node[black,midway,yshift=-0.6cm]{CSI collection};
    \draw[->] (3.2,-1.2) -- (3.8, -1.2);
    \filldraw[draw=black,fill=yellow] (\cyc,-1.4) rectangle +(2.8,0.5) node[pos=0.5] {BCA/GBA};
    \draw[->] (7,-1.2) -| (8,-0.7);
    \node at (10.1,-1) {New Resource Allocation};
    \draw[<->, color = blue] (4,-0.4) -- (6.8,-0.4) node[below, midway] {$\omega_c$};
 \end{tikzpicture}
 \caption{Example of pRUs allocation. Here, each cycle has $T=8$ slots, and $\eta=0.25$, meaning that $M=2$ devices can send a pilot over each channel at a given cycle in a round robin fashion, as shown with the numbers, using the pRUs (reported in colors). The computational time $\omega_c$ is lower than a cycle duration, so $W=2$ and the information collected at cycle $m$ is used for the allocation of the data RUs (in white) on cycle $m+2$.}
 \label{fig:overRUs}
\end{figure*}

Since the age of the CSI plays a pivotal role in the allocation efficiency, the overall performance of the system is determined by two quantities:
\begin{enumerate}
 \item the fraction $\eta$ of resources reserved for pilot transmission;
 \item the computational delay $W$ necessary to update the channel allocation.
\end{enumerate}
Let us analyze them separately.

As to the first element, we assume that on each channel $c$, $M = \eta T$ RUs are reserved for pilot, with $\eta\in(0,1)$. The devices are informed of which RUs are \emph{pilot RUs} (pRUs), which can be fixed or change across cycles, and exploit them in a round robin fashion.
At each cycle, $M$ devices are allowed to send a pilot: each of them transmits the pilot signal on one pRU per channel. This is meant to let the AP have a complete update of the entire CSI of a given device over all the channels in the same cycle, but other choices are feasible as well.

When GBA computes the resource allocation, the pRUs are taken into account, and the modified version of the algorithm, described in Section \ref{sec:GBA}, ensures that they are not assigned to any device.

In an idealized setup, where the allocation is computed instantaneously, the channel information collected during a cycle would be immediately used to derive the resource allocation in the subsequent cycle.
However, in practical scenarios, some time $\omega_c$ is needed for any allocation algorithm to run, and this time is not negligible in the analyzed scenario, where the cycle duration $\nu$ can be as low as a few milliseconds. Let us define the computational delay $W=\lceil \omega_c/\nu\rceil+1$. This value gives an indication about the time needed to exploit the collected CSI\footnote{To be more precise, we should consider both the computational time $\omega_c$ of the algorithm and the allocation information broadcast time $\omega_b$. While the former strongly depends on the characteristics of the adopted algorithm, the latter can be assumed fixed and performed over a robust downlink control channel. We consider it negligible in this study, but choosing a fixed positive value for $\omega_b$ would not significantly alter the proposed approach.}.
In fact, this means that the channel information collected during cycle $m$ is used for the allocation that will be adopted in cycle $m+W$. For example, if $\omega_c<\nu$, making $W=2$, at the end of cycle $m$ the AP starts computing the new allocation, and completes it before the end of the subsequent cycle $m+1$. Hence, the new allocation becomes effective in cycle $m+2$.
A sketch of the CSI collection functioning is illustrated in Figure~\ref{fig:overRUs}.

When computing the allocation, we assume that the computational delay is known. Hence, the system is aware of when the information will be used, and can therefore select the correct channel fading distribution. If, for example, the CSI of a given device $D_i$ has been collected in the present cycle, and $W=2$, the number of RUs required for $D_i$ on channel $c$ is found from (\ref{deffires}) with $t=2$.

Notice that the CSI age is not the same for all the devices. Looking at Figure 4, and assuming that only 6 devices are deployed, in cycle $m+2$ the CSI age of users $D_1$ and $D_2$ is equal to 2 cycles (since it was collected during cycle $m$), but it is instead equal to 3 for users $D_5$ and $D_6$ (collected in cycle $m-1$) and to 4 for users $D_3$ and $D_4$ (collected in cycle $m-2$).

\section{Dynamic Pilot Transmission Selection}
\label{sec:dynpilot}
A relevant issue of the considered allocation algorithm is fairness. Due to the impact of path loss, devices located farther from the AP experience worse channel conditions, and thus need, on average, a higher amount of RUs.
Hence, in order to allocate the highest amount of users, GBA first selects the closest ones, which can be accommodated with higher spectrum efficiency. This leaves small room for the farthest ones, and a fairness issue arises.

In this section, we propose a different scheme to organize the pilot transmissions, pursuing a fairer resource utilization while not sacrificing efficiency. The main idea is that devices located farther from the AP should send pilots more often than the closer ones. An updated CSI at the AP can consistently reduce the amount of needed RUs, and this is far more relevant for distant devices.

\subsection{Best Improvement Pilot Allocation}
Simple strategies for more efficient allocation of pRUs based only on topology can be envisioned. For instance, only devices farther than a threshold distance might be allowed to utilize the pRUs in a round robin fashion.
The strength of this approach is that it does not require additional information, since the scheme shows a periodic, deterministic allocation. However, the drawback is that devices deployed close to the threshold distance would suffer from a severely increased need for RUs.
Furthermore, topology alone cannot fully identify the devices that need more resources, since fading also plays a pivotal role. On the other hand, transmitters with good channel conditions may wait longer for a new CSI update than devices experiencing harsh conditions.
Since channel conditions are actually partially known at the AP, due to previously received pilots and to channel time correlation, we can design a better, dynamic strategy that leverages this information.

While keeping the fraction $\eta$ of pRUs fixed, the devices allowed to send pilot over these pRUs are now determined in a different way.
The choice of a given subset of devices to send their pilot affects system performance, and also its future evolution. From a theoretical perspective, one may develop a Markov Decision Process to determine the optimal pRUs allocation policy.
The states would be defined as a set of $N$ couples ($\mathbf{z}_i$, $u_i$), being $\mathbf{z}_i$ the vector of the last measured $C$ channel coefficients for device $D_i$ and $u_i$ the age of this information (number of cycles since the last pilot sent by $D_i$). An action would correspond to the selection of a subset of the $N$ devices for pilot transmission, which would change the state of the system accordingly.
Finally, the reward should be defined as a function of the overall amount of RUs required for data transmission after the CSI update, which is also a function of the new system state.
The well known Bellman equation expresses the optimal policy of the MDP. Unfortunately, even if the channel measurements are quantized, the overall number of states is huge, and this optimal approach is clearly not scalable for real time execution.

We resort instead to a heuristic suboptimal approach, that can quickly select the devices required to send pilot in each cycle. The rationale behind this approach is to choose the devices that can get the maximum advantage from a CSI updating.
The input for the algorithm, which is executed at the end of each cycle, is the set of $N$ couples ($\mathbf{z}_i$, $u_i$) as before, where $u_i=0$ if $D_i$ transmitted a pilot in the cycle that has just ended.
For device $D_i$ and channel $c$, the gain obtained through a pilot transmission is expressed as
\begin{equation}
 \phi_{ic} = \mathcal{F}(c,i,\rho,u_i+1|\mathbf{z}_i(c)) - \mathcal{F}(c,i,\rho,W|x_{ic}).
 \label{chagain}
\end{equation}
The former term is the number of RUs needed if the currently stored CSI is exploited. In fact, at the end of the next cycle, the CSI age of a device that does not transmit a pilot simply increases by 1. The latter is the number of RUs required if a new measurement $x_{ic}$ is collected through a pilot transmission in the incoming cycle. In this case, the CSI age is set to $W$, which takes into account the number of cycles still needed before the data resource allocation based on this new CSI will become effective.

The gain expressed in (\ref{chagain}) has two main drawbacks. Firstly, the measured coefficient $x_{ic}$ is not known a priori. Secondly, the gain depends on the channel, but we do not know in advance which channel will be allocated to the device for transmission.
To tackle the first point, we can rely on the expected value $\mathbb{E}[x_{ic}]$, which can be computed through the distribution given in Lemma~\ref{lem:corrchan} exploting the current CSI value $\mathbf{z}_i(c)$ and its age $u_i$.
As to the second issue, simulations actually show that GBA on average assigns each device to its best channel around 90\% of the times, and we can thus focus our attention only on it.
It is then reasonable to estimate the gain for device $D_i$ by
\begin{equation}
 \phi_i = \min_{c}\mathcal{F}(c,i,\rho,u_i+1|\mathbf{z}_i(c)) - \min_c\mathcal{F}(c,i,\rho,W|\bar{x}_{ic}),
 \label{gain}
\end{equation}
with
\begin{equation}
 \bar x_{ic} = \mathbb{E}[x_{ic}|\mathbf{z}_i(c)] = \int_0^{+\infty}xf_{u_i}(x|\mathbf{z}_i(c))\de x,
\end{equation}
and $f_{u_i}(x|\mathbf{z}_i(c))$ defined in (\ref{corrpdf}).
Once the gain is computed for all the $N$ devices, the pRUs are assigned to the $\eta N$ devices with the highest gain.

The proposed algorithm is inherently dynamic, since it tunes the frequency of pilot transmission based on the topology and on the channel conditions of each device. From the implementation point of view, the computation of (\ref{gain}) can be easily done by quantizing the fading coefficients.
Given the channel time correlation $\gamma$, a table can then be built offline, where each row corresponds to a quantized value of the previously measured fading coefficient, and each column to the information age in terms of cycles. Each element then contains the desired fading expected value for any combination of these two terms.
The overall complexity is $\mathcal{O}(NC)$, in order to find the gains of all the nodes, plus $\mathcal{O}(N\log N)$ to sort the obtained gains and select the devices that must transmit, which is lower than that of GBA~\cite{OurIoT}. In addition, since pRUs are known and fixed, it can be executed in parallel during the cycle, as the pilots are received, thus actually requiring neglibile time to be performed.

\section{Results}
\label{sec:results}
We simulated a network scenario with the parameters setup reported in Table \ref{tab:Sympara}. For each value of $\eta$ and $W$, we ran 10 simulations over random topologies, each lasting for 30 subsequent cycles. For each one, we then computed the number of served devices averaged over the second half of the simulation time (in order to avoid transient effects). The obtained value was then averaged again over all the tested topologies.
\begin{table}
 \centering
 \begin{tabular}{cc}
  \hline  \hline
  Parameter & Value \\
  \hline
  Deployment area radius $L$ & 60 m \\
  Channel correlation factor $\gamma$ & 0.95 \\
  Transmit SNR $\Gamma_T$ & 100 dB \\
  Path loss exponent $\alpha$ & 3 \\
  Packet size $\ell$ & 100 bit \\
  Slot duration $\tau$ & 0.144 ms \\
  Cycle duration $T$ & 50 slot \\
  Maximum delay $\Delta$ & 25 slot \\
  Channel bandwidth $B$ & 180 kHz \\
  Number of channels $C$ & 5 \\
  Maximum Interference factor $Y_M$ & 4 \\
  Target Reliability $\rho$ & 0.99999 \\
  \hline
 \end{tabular}
 \caption{Parameters for the symmetric scenario.}
 \label{tab:Sympara}
 \vspace{-0.8cm}
\end{table}
The spacing of the subcarriers is 15 kHz, and each OFDM symbol spreads over $n_c=12$ adjacent subcarriers, which form a channel of bandwidth 180 kHz. Each OFDM symbol duration is about 71.4 $\mu$s, and a RU is a sequence of 2 consecutive symbols, thus lasting $\tau=0.144$ ms.
As to the correlation factor $\gamma$, we set it by using the Jakes time correlation model~\cite{Jakcha}, so as $\gamma = J_0(2\pi f_D t_c)$, being $f_D$ the Doppler frequency, $t_c$ the channel instantiation interval, and $J_0(\cdot)$ the Bessel function of order 0. Frequency $f_D$ is obtained as $v_if_c/\bar c$, where $v_i$ is the speed of the device, $f_c$ is the carrier frequency and $\bar c$ is the speed of light.
By setting $f_c = 800$ MHz~\cite{ChaIIoT} and $t_c=\nu$, a speed $v_i=3$ km/h gives $\gamma=0.97$. Higher and lower values account for more static or more dynamic scenarios.
As to the pilot signaling, for any value of $\eta$, the $M$ pRUs were placed randomly and independently for each channel, since placing them all in a given sub-interval of the cycle would impair the devices with data packet issue time falling in or just before that interval.

In this section, we also compare the performance attained by GBA with that of a simpler benchmark allocation algorithm, presented in details in~\cite{OurIoT}, called BCA (Best Channel Allocation). BCA is a simple greedy heuristic that runs in quasi-linear time with $N$. After sorting the devices according to their data packet issue time, it checks them one at a time.
For each device $D_i$, based on the value $\mathcal{F}(c,i,\rho,z|t)$, and considering the pRUs and the devices already allocated on each channel, it finds the channel $c$ which grants the earliest transmission completion time.
If the transmission on channel $c$ can be completed within slot $t_i+\Delta$, hence matching the delay constraint, the corresponding RUs are assigned to $D_i$. Otherwise, device $D_i$ is excluded from the set of transmitters for this allocation round.
BCA has a much lower computational time, especially in dense networks, but its approach is suboptimal, since it does not give priority to devices with good channel conditions.

\begin{figure}
 \begin{center}    
 \includegraphics[width=\figw]{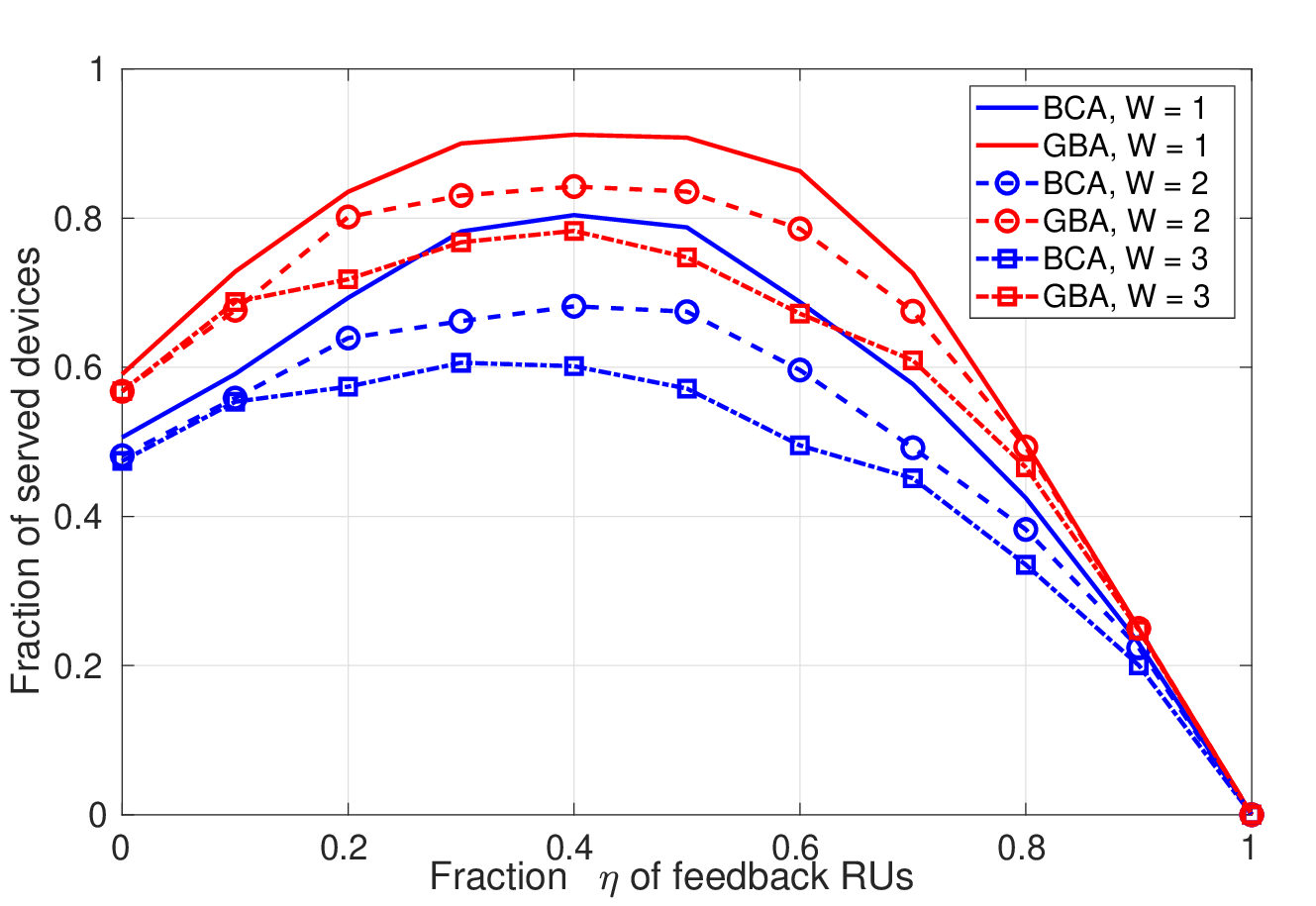}
 \caption{\small Fraction of served devices as a function of the fraction $\eta$ of pilot RUs, for various values of the computational delay $W$. Here, $N=100$, $\gamma=0.95$, $T=50$, $C=5$, $\Delta = 25$.}
 \label{fig:FB_fra_vs_res}
 \vspace{-0.4cm}
 \end{center}
\end{figure}

We observe the impact of both $\eta$ and $W$ in Figure~\ref{fig:FB_fra_vs_res}, where the average fraction of served devices is plotted against the fraction $\eta$ of RUs reserved for pilot transmission, for different values of $W$.
For $\eta=0$, meaning that no CSI is available at all, the performance is relatively low, due to the high uncertainty about the channel conditions. As $\eta$ increases, so does the number $M$ of devices whose CSI is updated at the end of each cycle. This in turns implies that the average value of $t$ in (\ref{deffires}) is reduced, and then also the average value of $\mathcal{F}(c,i,\rho,t|z)$ decreases.
As a result, more communications can be allocated, at the cost of having fewer RUs for data transmission.
When $\eta$ grows beyond 0.5, the advantages of a fresh CSI are overcome by the lack of resources, and the number of allocated devices quickly falls. Clearly, for $\eta=1$, there are no RUs left for data transmissions, and no device can transmit any more.

In the same figure, we can also assess the effect of $W$. The case $W=1$ is the idealized scenario, where the CSI collected in a cycle is immediately exploited in the allocation of the subsequent cycle. As $W$ becomes higher, the information collected is used later, with less effectiveness. As a consequence, the system attains a lower performance, with a loss of almost 10\%, when $\eta=0.5$, as $W$ grows from 1 to 2 and from 2 to 3.
As to the comparison between the resource allocation algorithms, GBA displays a clear advantage over BCA, over all ranges of $\eta$ and $W$. However, the value of $W$ needed to execute GBA might be larger than that needed for BCA due to its higher complexity, an issue that is addressed later on in this section. 

The channel correlation coefficient $\gamma$ strongly influences the effectiveness of CSI. A higher $\gamma$ reduces the impact of the CSI age, since the channel varies more slowly.
\begin{figure}
 \begin{center}    
 \includegraphics[width=\figw]{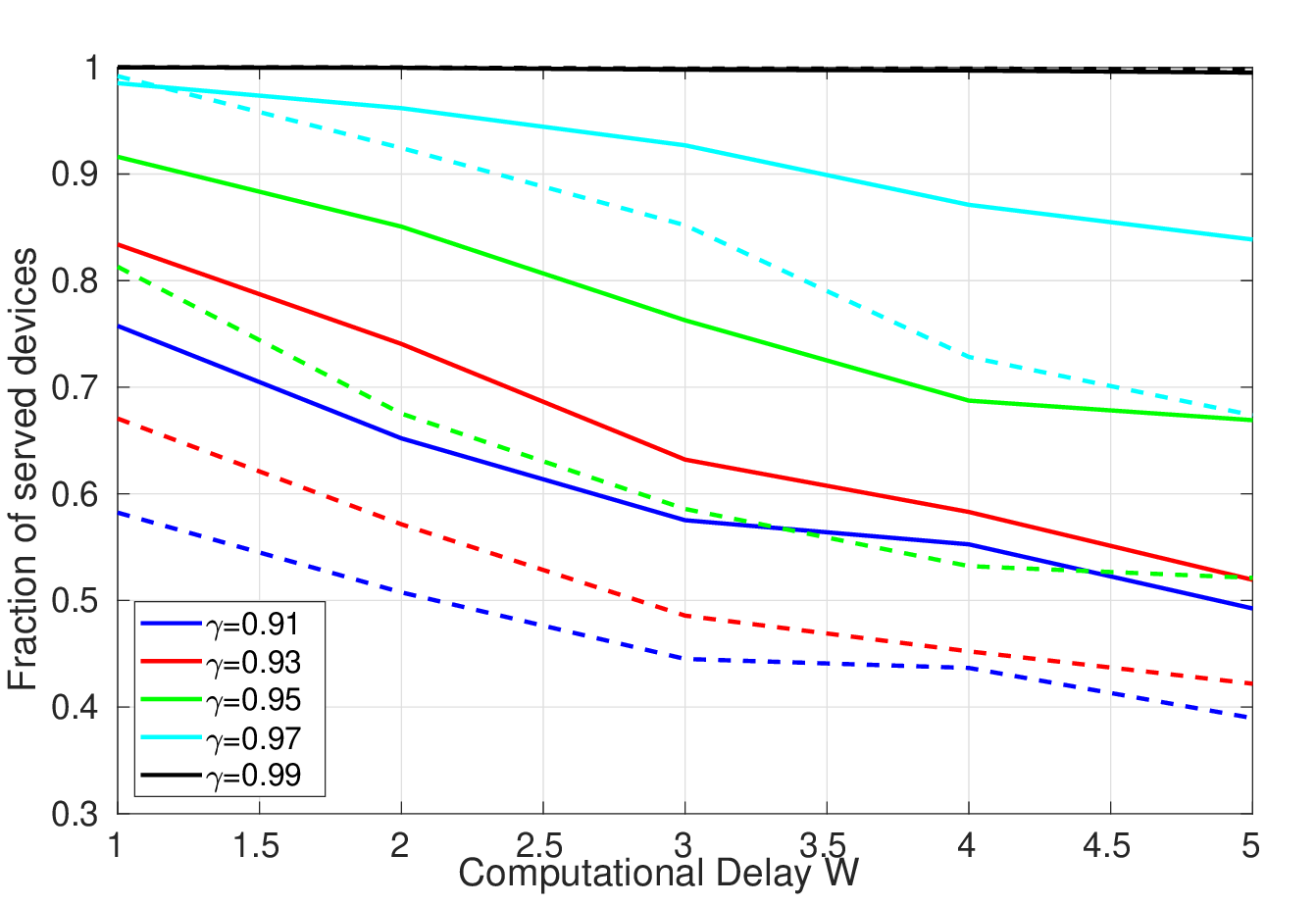}
 \caption{\small Fraction of served devices as a function of the computational delay $W$, when $\eta=0.4$, for different values of the channel correlation coefficient $\gamma$. Here, $N=100$, $T=50$, $C=5$, $\Delta = 25$. Solid lines are for GBA, while dashed lines are for BCA.}
 \label{fig:FB_fra_vs_W}
 \vspace{-0.4cm}
 \end{center}
\end{figure}
We analyze this aspect in Figure~\ref{fig:FB_fra_vs_W}, where we fix the fraction $\eta$ of pRUs to 0.4 and observe the fraction of served devices as a function of the computational delay $W$ for different correlation coefficients.
As expected, the performance decreases as $W$ becomes higher, but the impact of $\gamma$ is clear.
For low values of $\gamma$, even the idealized case of $W=1$ gives poor performance. In this case, the CSI quickly becomes outdated, and we get close to the scenario with no CSI at all. Conversely, for $\gamma>0.95$ in the idealized scenario all the devices can be always served.
Even more, for $\gamma=0.99$, the computational delay $W$ has little effect on the performance, which remains almost optimal also for high values of $W$.
In this case, the very high correlation makes the CSI age less relevant. This makes the average number of RUs needed per device very low, which allows the allocation of almost all the users on the fraction of resources reserved to data.
Looking at the result from a different point of view, slow allocation algorithms can be effectively applied when the channel correlation is very high. However, as $\gamma$ decreases, only fast algorithms, capable of keeping $W$ low, must be adopted in order to take advantage of the channel correlation.

\begin{figure}
 \begin{center}    
 \includegraphics[width=\figw]{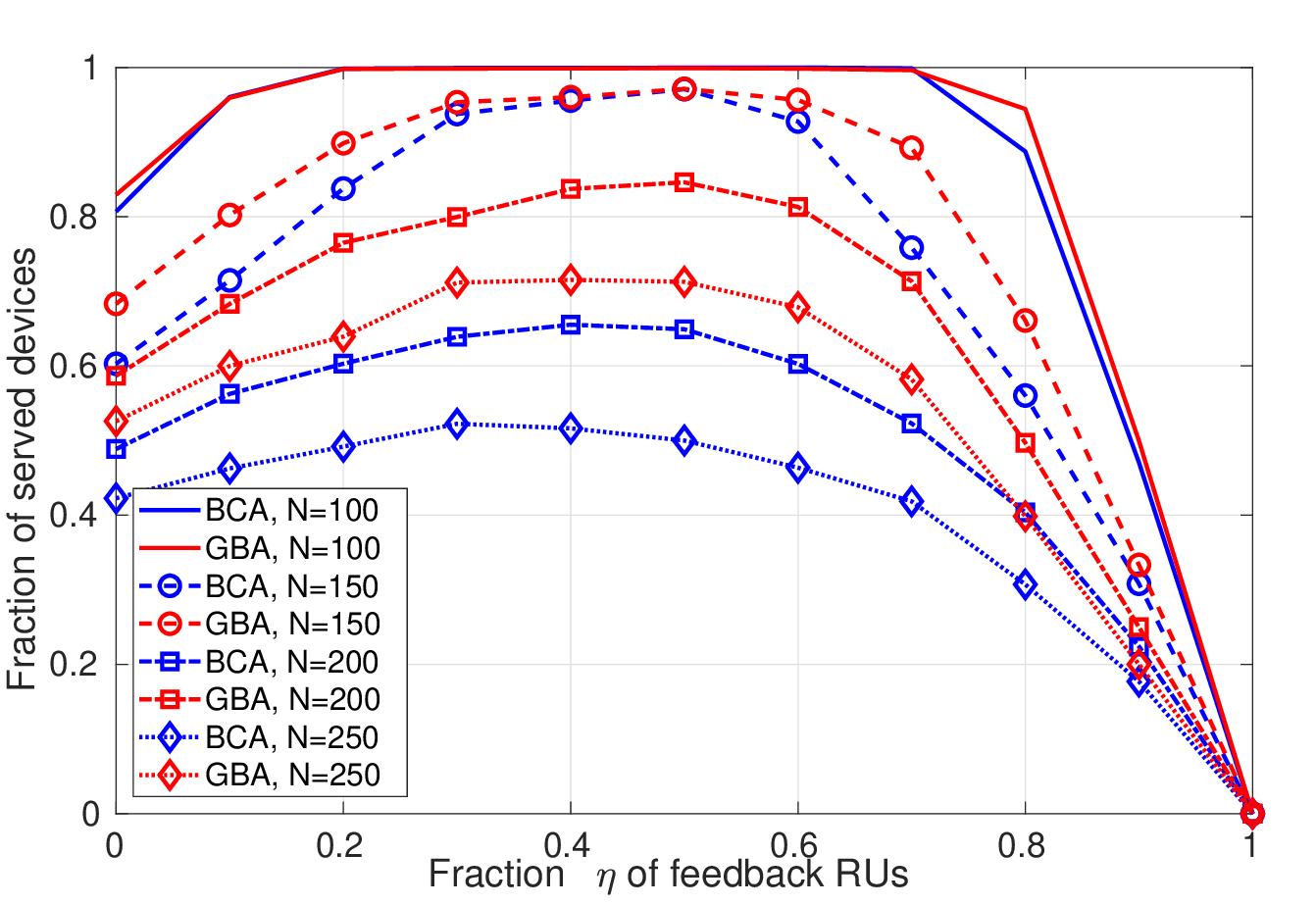}
 \caption{\small Fraction of served resources as a function of the fraction $\eta$ of pilot RUs, for various numbers of nodes $N$. Here, $\gamma=0.95$, $T=50$, $C=10$, $\Delta = 25$ slot.}
 \label{fig:FB_C10_fra_vs_res_varN}
 \vspace{-0.4cm}
 \end{center}
\end{figure}

Next, we move to another key issue, that is, the algorithm scalability. While BCA has a computational complexity which is quasi-linear in both $N$ and $C$, GBA performs better, but at the cost of a higher complexity, which was proved in~\cite{OurIoT} to be $\mathcal{O}((N+C)^2N^2C)$ in the worst case. Henceforth, as the number $N$ of devices grows, the advantages of the better performance might be impaired by the disadvantage of a higher computational delay $W$.
In order to test this aspect, we move to a more crowded scenario with $C=10$ channels, and double the number of devices up to $N=200$. In Figure~\ref{fig:FB_C10_fra_vs_res_varN} we plot again the fraction of served devices as a function of $\eta$ in this denser scenario. The general trend is the same as in Figure~\ref{fig:FB_fra_vs_res}, with an optimal $\eta$ again around 0.4 (although it tends to slightly decrease when $N$ grows, in order to leave room for the higher amount of data).

\begin{figure}
 \begin{center}    
 \includegraphics[width=\figw]{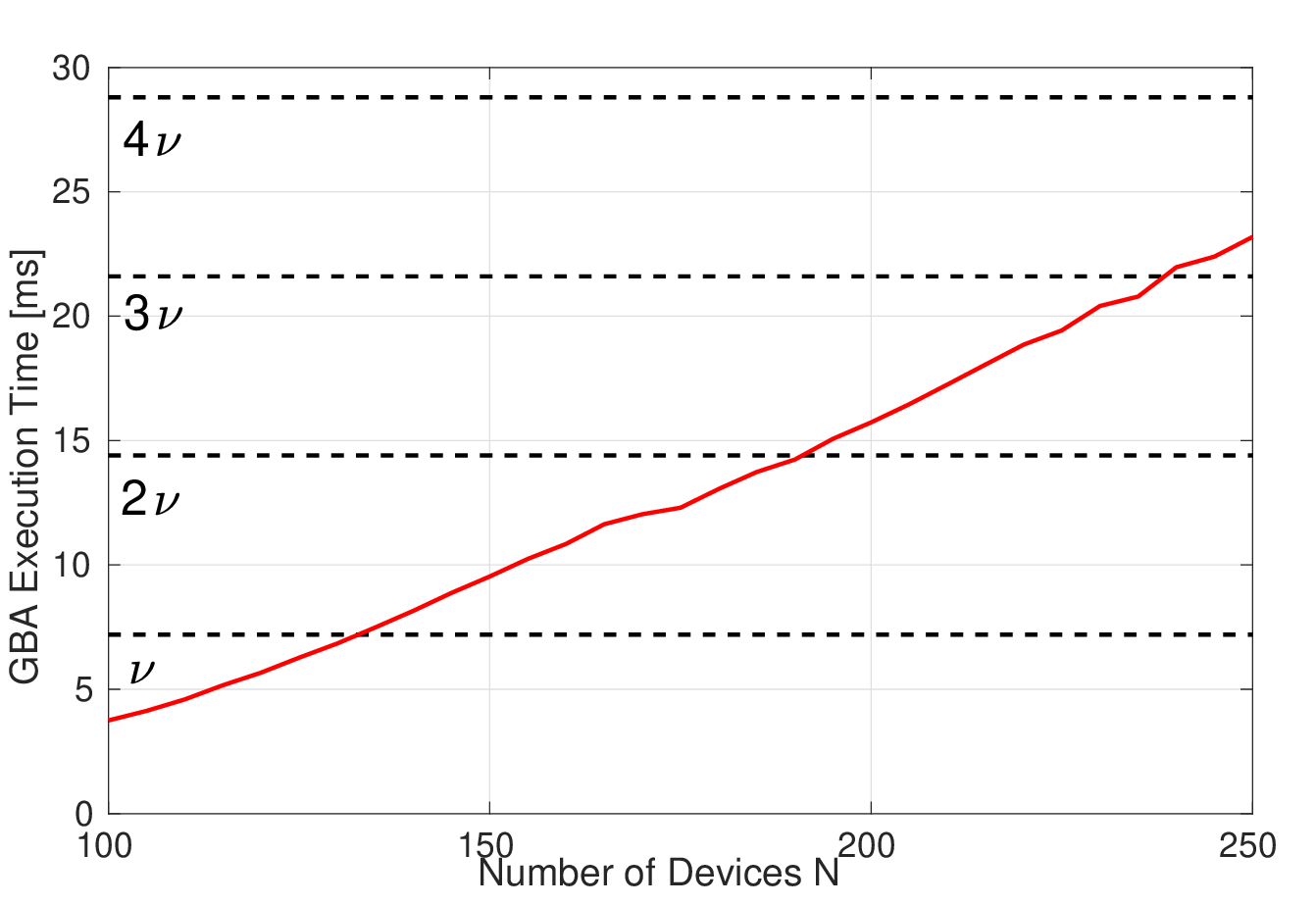}
 \caption{\small GBA running time on a commercial laptop, as a function of $N$. Here, $\gamma=0.95$, $T=50$, $C=10$, $\Delta = 25$.}
 \label{fig:FB_exetime_vs_N}
 \vspace{-0.4cm}
 \end{center}
\end{figure}

When running the simulations that led to the results in Figure~\ref{fig:FB_C10_fra_vs_res_varN}, we also measured and averaged the algorithm execution time for both BCA and GBA. While the former was always completed in less than 1 ms, the execution time of GBA as a function of the number $N$ of devices is plotted in Figure~\ref{fig:FB_exetime_vs_N}. The algorithm was run over a commercial laptop, and was implemented in C++. 

As expected, the execution time grows with $N$, and increases of almost a factor 4 when $N$ doubles from 100 to 200 devices. In order to assess the impact of the execution time on the performance, we need to compare it to the cycle time $\nu = T\tau$, which in the considered scenario is equal to 7.2 ms.
Hence, we also report the values of $\tau$ and its multiples in the figure, which highlights the number of cycles required for the algorithm to run. One cycle is enough for $N$ up to around 130 devices. In this interval, the computational delay $W$ is hence equal to 2. Two cycles are needed for $N$ up to 190 ($W=3$) and three for $N$ up to 235 ($W=4$).

By using these results, we can fairly compare BCA and GBA with the execution time also taken into account by properly varying $W$. For BCA, it can be kept to 2 for all the values of $N$, while we change it from 2 to 5, according to Figure~\ref{fig:FB_exetime_vs_N}, for GBA.

\begin{figure}
 \begin{center}    
 \includegraphics[width=\figw]{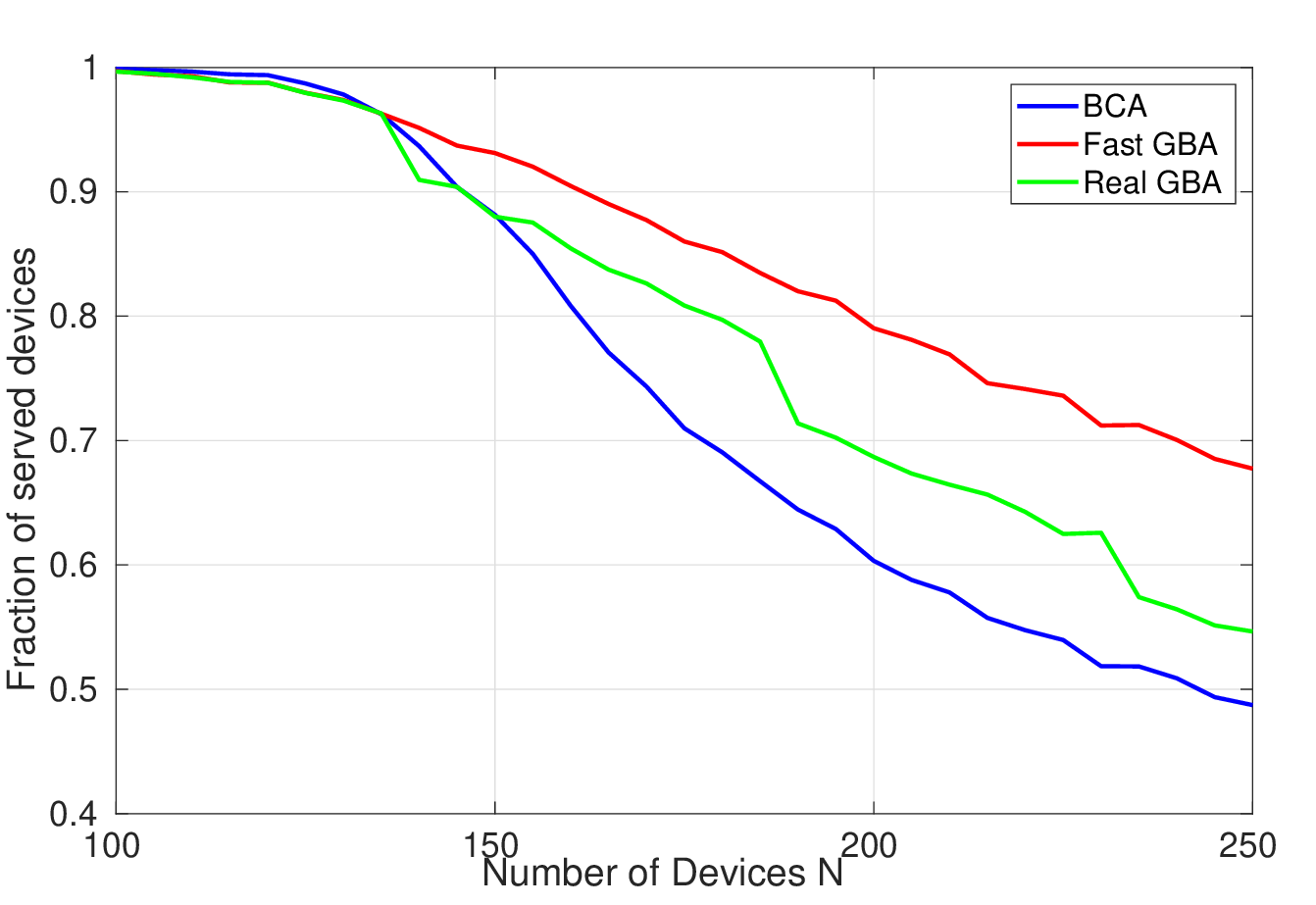}
 \caption{\small Fraction of served devices as a function of the network size $N$ taking into account the computation time.}
 \label{fig:FB_fra_vs_N}
 \vspace{-0.4cm}
 \end{center}
\end{figure}
We illustrate the obtained result in Figure~\ref{fig:FB_fra_vs_N}, where we compare BCA with an indealized, fast GBA (keeping $W=2$, as BCA) and a real GBA (with $W$ dependent on the computation time).
The increased computational time affects the fraction of served devices, which lowers by about 13\% for $N=250$ devices. Nevertheless, even in this computational intensive scenario, GBA still performs significantly better than BCA: its polynomial complexity contains the execution time within values low enough to still outperform the faster but suboptimal BCA. For $N=200$, GBA is able to serve 14\% more devices than BCA, and for $N=250$ the improvement is still around 12\%.

We recall that the reported results were obtained using a standard, non sophisticated implementation of GBA. The computational time gap between BCA and GBA could be reduced by improving it, e.g., using multi-thread computation. This way, the actual performance of GBA might get even closer to the idealized \emph{Fast GBA} line in Figure~\ref{fig:FB_fra_vs_N}.

Finally, we assess the impact of the dynamic pilot allocation scheme outlined in Section~\ref{sec:dynpilot} on GBA. We name as \emph{dynamic} GBA (d-GBA) the overall resulting protocol. We focus on a moderately dense scenario, with $N=150$ devices, and $C=7$ channels, with a correlation factor $\gamma=0.97$.
\begin{figure}
 \begin{center}    
 \includegraphics[width=\figw]{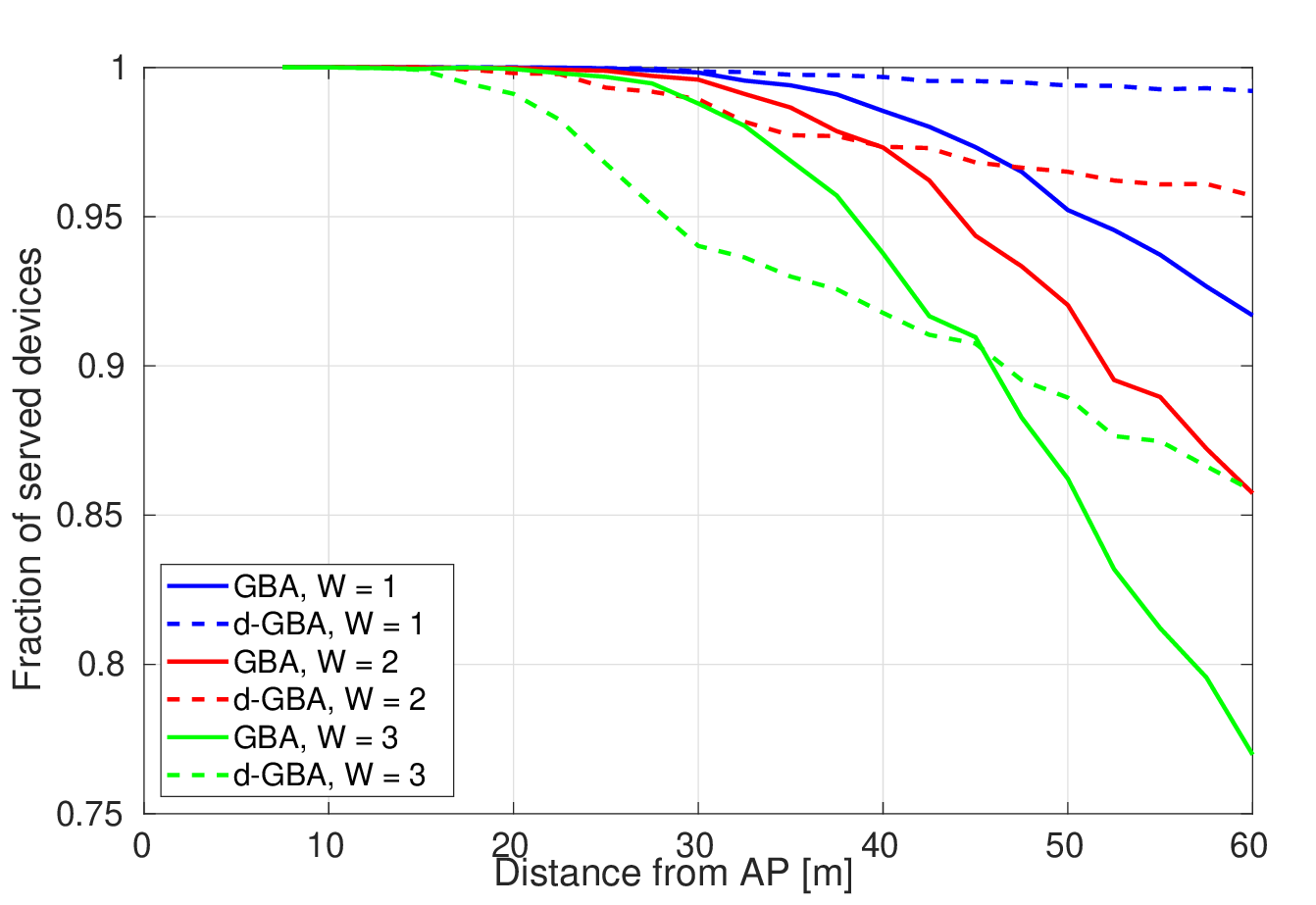}
 \caption{\small Fraction of served devices as a function of the distance from the AP. Here $\eta= 0.4$.}
 \label{fig:FB_fra_vs_dist_pilot}
 \vspace{-0.4cm}
 \end{center}
\end{figure}
In Figure~\ref{fig:FB_fra_vs_dist_pilot}, we report the transmission probability for a device as a function of its distance from the AP. As expected, basic GBA with a round robin pilot selection serves all the devices in the inner part of the network (up to 30 m).
However, the fraction of served devices starts to fall beyond this value, with a rate dependent on the computational delay $W$. Only 85\% and 77\% of the edge devices are actually served, for the realistic cases of $W=2$ and $W=3$.
Our proposed dynamic pilot transmission selection scheme can consistently mitigate this impairment, at the cost of slightly lowering the fraction of served devices at mid-distance. Looking at the performance of d-GBA, in fact, the percentage of served edge devices can be increased up to 96\% and 85\% when $W=2$ and $W=3$, respectively, thus granting a considerable benefit in terms of fairness.

\begin{figure}
 \begin{center}    
 \includegraphics[width=\figw]{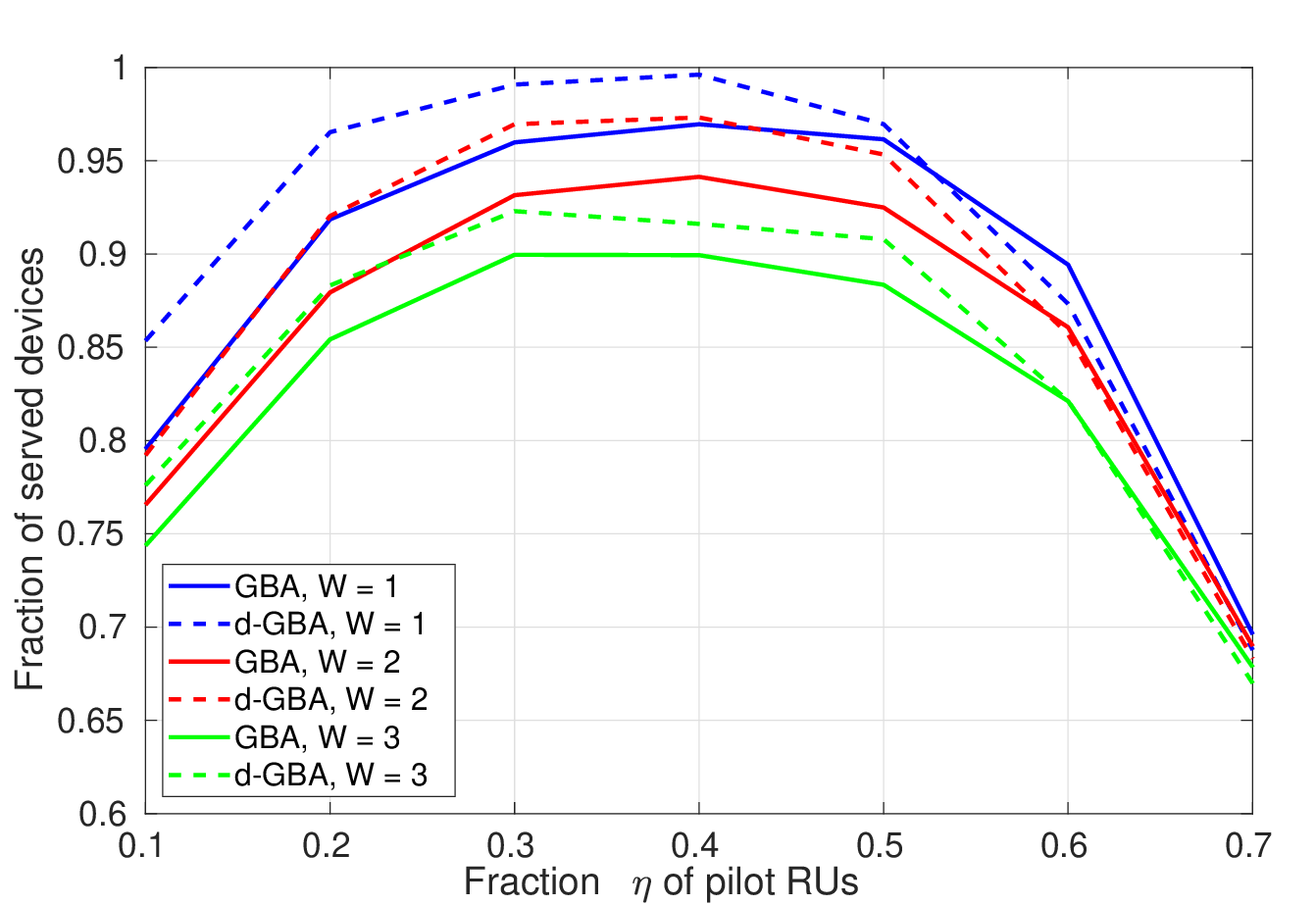}
 \caption{\small Fraction of served devices as a function of the fraction $\eta$ of pRUs, when the dynamic pilot transmission selection is adopted.}
 \label{fig:FB_fra_vs_eta_pilot}
 \vspace{-0.4cm}
 \end{center}
\end{figure}
Notably, this fairness improvement does not come at the expense of the spectrum efficiency, as highlighted in Figure~\ref{fig:FB_fra_vs_eta_pilot}. Here, we plot the overall fraction of served devices as a function of the fraction $\eta$ of pRUs, with all the other parameters unchanged.
The dynamic selection of the devices sending pilots at every cycle can better balance the spectrum requirements of all the users, thus allowing d-GBA to allocate more transmissions. Almost the entire set of devices can be served, for $\eta=0.4$, in the idealized case $W=1$, but a performance boost is achieved for higher values of $W$ as well.
Indeed, the number of served devices is increased along the entire range of $\eta$ values, of as much as nearly 5\% for the best performing $\eta$ values.

\section{Conclusions}
\label{sec:conclu}
In this paper, we investigated the issue of efficiently collecting and exploiting CSI in order to improve spectrum resource allocation in an industrial IoT scenario leveraging URLLC with hundreds of devices and periodic traffic pattern.
We considered time-correlated channels, where CSI relevance is kept for subsequent transmission cycles. We analyzed the tradeoff between a more frequent pilot signaling, able to reduce the CSI average age, and a lower overhead option, where more room is left for data transmission.
We investigated the tradeoff between computational complexity and performance, showing that a sophisticated graph-based allocation algorithm, despite its longer computational time, is able to provide a much better resource allocation solution than that offered by a greedy heuristic.
In addition, we designed and added a dynamic pilot transmissions allocation, able to adaptively tune the CSI age based on topology and channel measurements, thus further boosting the performance in terms of both fairness and spectrum efficiency.

\appendices
\section{Proof of Lemma \ref{lem:corrchan}}
\label{app:proof2}

To prove the Lemma, it is enough to derive the pdf of the random variable $|X|^2$, with $X$ defined as
\begin{equation}
 X = a_0y + \sum_{i=1}^na_iZ_i,
 \label{defX}
\end{equation}
where $y\in\mathbb{C}$, $a_i\in\mathbb{R}^+$ and the $Z_i$'s are independent complex Gaussian random variable with zero mean and unit variance. For ease of notation, we call the real and imaginary parts of the involved complex quantities as $y_R = \Re[y]$, $y_I = \Im[y]$, $Z_{iR} = \Re[Z_i]$ and $Z_{iI}=\Im[Z_i]$.
Each complex Gaussian random variable $Z_i$ can be decomposed as $(Z_{iR} + jZ_{iI})/\sqrt{2}$, where $j$ is the imaginary unit, while $Z_{iR},Z_{iI}\sim\mc{N}(0,1)$.
We can hence rewrite (\ref{defX}) as
\begin{eqnarray}
 X & = & a_0y_R + ja_0y_I + \sum_{i=1}^n\frac{a_i}{\sqrt{2}}Z_{iR} + j\sum_{i=1}^n\frac{a_i}{\sqrt{2}}Z_{iR} \nonumber \\
 & = & \sum_{i=1}^n\left(\frac{a_0y_R}{n} + \frac{a_i}{\sqrt{2}}Z_{iR}\right)\! +\! j\sum_{i=1}^n\left(\frac{a_0y_I}{n} + \frac{a_i}{\sqrt{2}}Z_{iI}\right).\nonumber\\
\end{eqnarray}
Therefore, the squared absolute value is
\begin{eqnarray}
 |X|^2 &\hspace{-0.3cm} = &\hspace{-0.25cm} \left|\sum_{i=1}^n\left(\frac{a_0y_R}{n}\! +\! \frac{a_i}{\sqrt{2}}Z_{iR}\right)\right|^2\!\!\! +\! \left|\sum_{i=1}^n\left(\frac{a_0y_I}{n}\! +\! \frac{a_i}{\sqrt{2}}Z_{iI}\right)\right|^2 \nonumber \\
 &\hspace{-0.3cm} = &\hspace{-0.25cm} \left(\frac{1}{2}\sum_{i=1}^na_i^2\right)\left[\left|\sum_{i=1}^n\left(\frac{a_0y_R}{n\sqrt{\sum_ia_i^2/2}} + \frac{a_iZ_{iR}}{\sqrt{\sum_ia_i^2}}\right)\right|^2\right. \!\! + \nonumber \\
 &\hspace{-0.3cm} &\hspace{-0.25cm} + \left.\left|\sum_{i=1}^n\left(\frac{a_0y_I}{n\sqrt{\sum_ia_i^2/2}} + \frac{a_iZ_{iI}}{\sqrt{\sum_ia_i^2}}\right)\right|^2\right]\nonumber\\
 & \hspace{-0.3cm}= &\hspace{-0.25cm} \left(\frac{1}{2}\sum_{i=1}^na_i^2\right)\left(|X_R|^2 + |X_I^2|\right),
 \label{derix2}
\end{eqnarray}
where $X_R$ and $X_I$ are sums of $n$ independent Gaussian random variables, and are therefore themselves Gaussian, with $X_R\sim\mc{N}(a_0y_R/\sqrt{\sum_ia_i^2/2},1)$ and $X_I\sim\mc{N}(a_0y_I/\sqrt{\sum_ia_i^2/2},1)$.
 
Let us define the random variable $Y = 2|X|^2/\sum_ia_i^2$. According to (\ref{derix2}), $Y$ is given by the sum of the squared absolute value of two independent Gaussian random variables of unitary variance, and is therefore distributed as a non-central $\chi$-square random variable with two degrees of freedom and parameter
\begin{equation}
 \lambda = \frac{2a_0^2y_R^2}{\sum_ia_i^2} + \frac{2a_0^2y_I^2}{\sum_ia_i^2} = \frac{2a_0^2}{\sum_ia_i^2}|y|^2.
\end{equation}
The Cumulative Distribution Function of $|X|^2$ can hence be found as
\begin{equation}
 F_{|X|^2}(x) = \mathbb{P}\left[|X|^2<x\right] = \mathbb{P}\!\left[Y\!<\frac{2x}{\sum_ia_i^2}\right] = F_Y\!\left(\frac{2x}{\sum_ia_i^2}\right)
\end{equation}
while the pdf is obtained by derivation as
\begin{eqnarray}
 f_{|X|^2}(x) &\!\! =\!\! & \frac{2}{\sum_ia_i^2}f_Y\left(\frac{2x}{\sum_ia_i^2}\right) \nonumber \\
 &\!\! = &\!\! \frac{1}{\sum_ia_i^2}e^{-(x+a_0|y|^2)/\sum_ia_i^2}I_0\left(\frac{2a_0}{\sum_ia_i^2}\sqrt{|y|^2x}\right),\nonumber \\
 \label{finexpre}
\end{eqnarray}
where we used the expression of the pdf of the non-central $\chi$-square distribution.
 
In our specific problem, by recursively applying the relationship in (\ref{corrchan}), we find
\begin{equation}
 h_{i,c,m+t} = \gamma^th_{i,c,m} + \sqrt{1-\gamma^2}\sum_{i=1}^t\gamma^{t-i}\xi_i,
\end{equation}
an expression equivalent to (\ref{defX}), where $y=h_{i,c,m}$, $a_0=\gamma^t$ and $a_i=\sqrt{1-\gamma^2}\gamma^{t-i}$. The distribution of $|h_{i,c,m+t}|^2$, conditioned on the value of $|h_{i,c,m}|^2=z$, is hence given by (\ref{finexpre}) by replacing $|y|^2$ with $z$, $a_0$ with $a=\gamma^t$ and $\sum_ia_i^2$ with $b=(1-\gamma^2)\sum_{i=0}^{t-1}\gamma^{2i}$, which completes the proof.

\bibliographystyle{IEEEtran}
\bibliography{IEEEabrv,biblio}

\newcommand{\noop}[1]{}
\begin{thebibliography}{10}
\providecommand{\url}[1]{#1}
\csname url@rmstyle\endcsname
\providecommand{\newblock}{\relax}
\providecommand{\bibinfo}[2]{#2}
\providecommand\BIBentrySTDinterwordspacing{\spaceskip=0pt\relax}
\providecommand\BIBentryALTinterwordstretchfactor{4}
\providecommand\BIBentryALTinterwordspacing{\spaceskip=\fontdimen2\font plus
\BIBentryALTinterwordstretchfactor\fontdimen3\font minus
  \fontdimen4\font\relax}
\providecommand\BIBforeignlanguage[2]{{%
\expandafter\ifx\csname l@#1\endcsname\relax
\typeout{** WARNING: IEEEtran.bst: No hyphenation pattern has been}%
\typeout{** loaded for the language `#1'. Using the pattern for}%
\typeout{** the default language instead.}%
\else
\language=\csname l@#1\endcsname
\fi
#2}}

\bibitem{Rew1in}
Y.~{Zhou}, L.~{Tian}, L.~{Liu}, and Y.~{Qi}, ``Fog computing enabled future
  mobile communication networks: A convergence of communication and
  computing,'' \emph{{IEEE} Commun. Mag.}, vol.~57, no.~5, pp. 20--27, May
  2019.

\bibitem{Rew2in}
Y.~Zhou, L.~Liu, L.~Wang, N.~Hui, X.~Cui, J.~Wu, Y.~Peng, Y.~Qi, and C.~Xing,
  ``Service-aware {6G}: An intelligent and open network based on the
  convergence of communication, computing and caching,'' \emph{Digital
  Communications and Networks}, vol.~6, no.~3, pp. 253--260, Aug. 2020.

\bibitem{OurIoT}
F.~Librino and P.~Santi, ``Resource allocation and sharing in {URLLC} for {IoT}
  applications using shareability graphs,'' \emph{IEEE Internet of Things J.},
  vol.~7, no.~10, pp. 10\,511 -- 10\,526, Oct. 2020.

\bibitem{M5}
S.~E. Elayoubi, P.~Brown, M.~Deghel, and A.~Galindo-Serrano, ``Radio resource
  allocation and retransmission schemes for {URLLC} over {5G} networks,''
  \emph{{IEEE} J. Select. Areas Commun.}, vol.~37, no.~4, pp. 896 -- 904, Apr.
  2019.

\bibitem{M23}
\BIBentryALTinterwordspacing
3GPP. (2017, Apr) \emph{Chairman's Notes}, document {R1-1704170, 3GPP: 3GPP TSG
  RAN WG1 Meeting 88bis}. [Online]. Available:
  \url{http://www.3gpp.org/ftp/TSG_RAN/WG1_RL1/TSGR1_88b/Report/}
\BIBentrySTDinterwordspacing

\bibitem{M3}
M.~Bennis, M.~Debbah, and H.~V. Poor, ``Ultrareliable and low-latency wireless
  communication: Tail, risk and scale,'' \emph{Proc. {IEEE}}, vol. 106, no.~10,
  pp. 1834 -- 1853, Oct. 2018.

\bibitem{M4}
P.~Popovski, \emph{et~al.}, ``Wireless access for ultra-reliable low-latency
  communication: Principles and building blocks,'' \emph{{IEEE} Network},
  vol.~32, no.~2, pp. 16 -- 23, Apr. 2018.

\bibitem{M6}
A.~Azari, M.~Ozger, and C.~Cavdar, ``Risk-aware resource allocation for
  {URLLC}: Challenges and strategies with machine learning,'' \emph{{IEEE}
  Commun. Mag.}, vol.~57, no.~3, pp. 42 -- 48, Mar. 2019.

\bibitem{M25}
N.~B. Khalifa, V.~Angilella, M.~Assaad, and M.~Debbah, ``Low-complexity channel
  allocation scheme for {URLLC} traffic,'' \emph{{IEEE} Trans. Commun.}, Sep.
  2020, {E}arly Access.

\bibitem{M27}
J.~{Cheng}, C.~{Shen}, and S.~{Xia}, ``Robust urllc packet scheduling of ofdm
  systems,'' in \emph{2020 IEEE Wireless Communications and Networking
  Conference (WCNC)}, 2020, pp. 1--6.

\bibitem{M1}
B.~Chang, L.~Zhang, L.~Li, G.~Zhao, and Z.~Chen, ``Optimizing resource
  allocation in {URLLC} for real-time wireless control systems,'' \emph{{IEEE}
  Trans. Veh. Technol.}, vol.~68, no.~9, pp. 8916 -- 8927, Sep. 2019.

\bibitem{M8}
W.~R. Ghanem, V.~Jamali, Y.~Sun, and R.~Schober, ``Resource allocation for
  multi-user downlink {URLLC-OFDMA} systems,'' in \emph{Proc. IEEE ICC}, Jun.
  2019.

\bibitem{M11}
C.~Sun, C.~She, C.~Yang, T.~Q.~S. Quek, Y.~Li, and B.~Vucetic, ``Optimizing
  resource allocation in the short blocklength regime for ultra-reliable and
  low-latency communications,'' \emph{{IEEE} Trans. Wireless Commun.}, vol.~18,
  no.~1, pp. 402 -- 415, Jan. 2019.

\bibitem{M13}
C.~She and C.~Y. amd T.~Q.~S.~Quek, ``Cross-layer optimization for
  ultra-reliable and low-latency radio access networks,'' \emph{{IEEE} Trans.
  Wireless Commun.}, vol.~17, no.~1, pp. 127 -- 141, Jan. 2018.

\bibitem{M15}
Z.~Chu, W.~Yu, P.~Xiao, F.~Zhou, N.~Al-Dhahir, A.~ul~Quddus, and R.~Tafazolli,
  ``Opportunistic spectrum sharing for {D2D}-based {URLLC},'' \emph{{IEEE}
  Trans. Veh. Technol.}, vol.~68, no.~9, pp. 8995 -- 9006, Sep. 2019.

\bibitem{ML1}
J.~{Li} and X.~{Zhang}, ``Deep reinforcement learning-based joint scheduling of
  {eMBB} and {URLLC} in {5G} networks,'' \emph{{IEEE} Wireless Commun. Lett.},
  vol.~9, no.~9, pp. 1543--1546, Sep. 2020.

\bibitem{ML2}
Q.~{Huang}, X.~{Xie}, H.~{Tang}, T.~{Hong}, M.~{Kadoch}, K.~K. {Nguyen}, and
  M.~{Cheriet}, ``Machine-learning-based cognitive spectrum assignment for {5G
  URLLC} applications,'' \emph{{IEEE} Network}, vol.~33, no.~4, pp. 30--35,
  Jul. 2019.

\bibitem{ML3}
H.~{Khan}, M.~M. {Butt}, S.~{Samarakoon}, P.~{Sehier}, and M.~{Bennis}, ``Deep
  learning assisted {CSI} estimation for joint {URLLC} and {eMBB} resource
  allocation,'' in \emph{2020 IEEE International Conference on Communications
  Workshops (ICC Workshops)}, 2020, pp. 1--6.

\bibitem{M26}
Y.~{Xie}, P.~{Ren}, and D.~{Xu}, ``Transmission performance optimization for
  {URLLC} with limited training and feedback overheads,'' \emph{IEEE Access},
  vol.~8, pp. 140\,467--140\,477, Jul. 2020.

\bibitem{M28}
J.~{Zeng}, T.~{Lv}, R.~P. {Liu}, X.~{Su}, N.~C. {Beaulieu}, and Y.~J. {Guo},
  ``Linear minimum error probability detection for massive {MU-MIMO} with
  imperfect {CSI} in {URLLC},'' \emph{{IEEE} Trans. Veh. Technol.}, vol.~68,
  no.~11, pp. 11\,384--11\,388, Nov 2019.

\bibitem{M29}
J.~{Zeng}, T.~{Lv}, R.~P. {Liu}, X.~{Su}, Y.~J. {Guo}, and N.~C. {Beaulieu},
  ``Enabling ultrareliable and low-latency communications under shadow fading
  by massive {MU-MIMO},'' \emph{IEEE Internet of Things J.}, vol.~7, no.~1, pp.
  234--246, Jan. 2020.

\bibitem{M30}
E.~J. {dos Santos}, R.~D. {Souza}, J.~L. {Rebelatto}, and H.~{Alves}, ``Network
  slicing for {URLLC} and {eMBB} with max-matching diversity channel
  allocation,'' \emph{{IEEE} Commun. Lett.}, vol.~24, no.~3, pp. 658--661, Mar.
  2020.

\bibitem{M31}
B.~{Bai}, W.~{Chen}, Z.~{Cao}, and K.~B. {Letaief}, ``Max-matching diversity in
  {OFDMA} systems,'' \emph{{IEEE} Trans. Commun.}, vol.~58, no.~4, pp.
  1161--1171, Mar. 2010.

\bibitem{hungarian}
D.~B. West, \emph{{Introduction to Graph Theory}}.\hskip 1em plus 0.5em minus
  0.4em\relax {Prentice Hall Upper Saddle River}, 2001.

\bibitem{New1}
M.~Medard, ``The effect upon channel capacity in wireless communications of
  perfect and imperfect knowledge of the channel,'' \emph{{IEEE} Trans. Inform.
  Theory}, vol.~46, no.~3, pp. 933 -- 946, May 2000.

\bibitem{M12}
W.~Yang, G.~Durisi, T.~Koch, and Y.~Polyanskiy, ``Quasi-static multiple-antenna
  fading channels at finite blocklength,'' \emph{{IEEE} Trans. Inform. Theory},
  vol.~60, no.~7, pp. 4232 -- 4265, Jul. 2014.

\bibitem{Zorzi}
B.~Makki, T.~Svensson, and M.~Zorzi, ``Finite block-length analysis of the
  incremental redundancy {HARQ},'' \emph{{IEEE} Wireless Commun. Lett.},
  vol.~3, no.~5, pp. 529 -- 532, Oct. 2014.

\bibitem{CSIerr}
G.~{Fodor}, P.~D. {Marco}, and M.~{Telek}, ``On minimizing the {MSE} in the
  presence of channel state information errors,'' \emph{{IEEE} Commun. Lett.},
  vol.~19, no.~9, pp. 1604--1607, Sep. 2015.

\bibitem{Jakcha}
H.~Kim and J.~Choi, ``A new design of polar-cap differential codebook for
  temporally and spatially correlated {MISO} channels,'' \emph{{IEEE} Trans.
  Wireless Commun.}, vol.~11, no.~2, pp. 703 -- 711, Feb. 2012.

\bibitem{ChaIIoT}
W.~Wang, S.~L. Capitaneanu, D.~Marinca, and E.-S. Lohan, ``Comparative analysis
  of channel models for industrial {IoT} wireless communication,'' \emph{IEEE
  Access}, vol.~7, pp. 91\,627 -- 91\,640, Jul. 2019.

\end{thebibliography}

\end{document}